\documentclass[a4paper,11pt]{article}
\usepackage{jcappub} % for details on the use of the package, please
\usepackage{mathtools}
\usepackage{hyperref}
\usepackage[T1]{fontenc} % if needed
\usepackage{subcaption}
\usepackage{amsmath} % For mathematical symbols and environments
\usepackage{amsfonts} % For mathematical fonts
\usepackage{amssymb} % For additional mathematical symbols
\usepackage{bm} % For bold math symbols

\def\HI{H{\sc i}\, }

\title{\boldmath Calibration requirements for Epoch of Reionization 21-cm signal observations - IV. Bias and variance with time and frequency correlated residual gains}

\def \iitbhu {Department of Physics, Indian Institute of Technology (Banaras Hindu University), Varanasi - 221005, India}
\def \iisc {Department of Physics, Indian Institute of Science, Bangalore 560012, India}
\def \iitm {Centre for Strings, Gravitation and Cosmology, Department of Physics, Indian Institute of Technology, Madras, Chennai 600036, India}
\def \knpg  {Department of Physics, K. N. Government P. G. College, Gyanpur, Bhadohi - 221304, India}

\author[a, 1]{Saikat Gayen  \note{Equal contribution}}
\author[b, 1]{Jais Kumar}
\author[a]{Prasun Dutta}
\author[c]{Khandakar Md Asif Elahi}
\author[c]{Samir Choudhuri}
\author[d]{Nirupam Roy}

\affiliation[a]{\iitbhu}
\affiliation[b]{\knpg}
\affiliation[c]{\iitm}
\affiliation[d]{\iisc}

\emailAdd{saikatgayen.rs.phy22@itbhu.ac.in}
\emailAdd{jaisk.rs.phy16@itbhu.ac.in}
\emailAdd{pdutta.phy@itbhu.ac.in}
\emailAdd{asifelahi999@gmail.com}
\emailAdd{samir@iitm.ac.in}
\emailAdd{nroy@iisc.ac.in }

\abstract{
Observation of multifrequency angular power spectrum of the redshifted 21-cm brightness temperature fluctuation from the neutral hydrogen holds the key to understand the structure formation and its evolution during the reionization and post-reionization era. A major challenge in observing the neutral hydrogen arises from presence of strong foreground signals in the frequency range of interest. Mitigating the direct effect of foregrounds are being addressed through various techniques in literature. An additional second order effect arises, in presence of foreground, with limited accuracy in time and frequency dependent gain calibrations. This manifests as the residual gain and bandpass error in the observed data, introduces bias and increases uncertainty in the estimates of 
multifrequency angular power spectrum. In this work, we present an analytic method to estimate the bias and excess uncertainty in the estimates of multifrequency angular power spectrum in presence of residual gain and bandpass errors. We use this framework to estimate the effect of these errors for detection of redshifted 21-cm emission from a redshift of $\sim 8$ with the upcoming SKA1-Low. Due to the high baseline density at the required range of angular multipoles, the SKA1-Low is found to be a tuned instrument for the redshifted 21-cm signal detection. We find that, there are scenario with residual gain and bandpass errors where there can be significant bias in these estimates. Certain foreground mitigation strategies,  is expected to reduce a part of the bias. The detailed study of different aspects of gain and bandpass errors and their relative effects are discussed. We find, with assumed models of gain and bandpass errors, signal detection is possible at this redshift with $128$ hours of observations. However, to achieve this one needs to have better calibration accuracy than present day interferometers. }


\begin{document}
\maketitle
\flushbottom

\section{Introduction}
\label{sec:introduction}

The cosmological 21-cm power spectrum serves as a valuable tool for studying the distribution of neutral hydrogen (\HI) on large scales across a wide range of redshifts, spanning from the Dark Ages to the Post-Reionization Era. Numerous studies have prescribed how measurements of the 21-cm power spectrum can be used to explore this cosmic evolution \citep{2005MNRAS.356.1519B, 2006AJ....131.1203F, 2019MNRAS.483.5480M, 2012RPPh...75h6901P, 2013ExA....36..235M}.
To probe the Epoch of Reionization (EoR), various ongoing and future experiments have been designed. These include the 
upgraded Giant Metrewave Radio Telescope (uGMRT) \citep{1991CuSc...60...95S, 2017CSci..113..707G}, Low-Frequency Array \citep[LOFAR;][]{2013A&A...556A...2V}, Murchison Widefield Array \citep[MWA;][]{2013PASA...30....7T, 2013PASA...30...31B}, the Donald C. Backer Precision Array for Probing the Epoch of Reionization \citep[PAPER;][]{2010AJ....139.1468P, 2015ApJ...809...61A},
the Hydrogen Epoch of Reionization Array \citep[HERA;][]{2017PASP..129d5001D}, the New Extension in Nançay
Upgrading loFAR \citep[NenuFAR;][]{2012sf2a.conf..687Z, 7136773}, the Square Kilometer Array \citep{2013ExA....36..235M, 2015aska.confE...1K} etc. These experiments aim to detect and characterize the 21-cm power spectrum during the cosmic dawn, EoR and post reionization era, shedding light on the early universe.
%Foregrounds and Removal
However, the redshifted 21-cm signal is faint, amid foreground emissions that are orders of magnitude stronger than the expected signal making it challenging to detect \citep[e.g.][]{1999A&A...345..380S, 2005ApJ...625..575S, 2008MNRAS.385.2166A, 2012MNRAS.426.3295G, Bernardi}. These foregrounds encompass various sources such as unresolved point sources, diffuse Galactic synchrotron emission, and free-free emission from our Galaxy and external galaxies. To overcome this issue, several techniques have been proposed.
One approach is the foreground subtraction technique, which involves subtracting a foreground model from the visibility data or the image. The residual data after foreground subtraction can then be used to detect the 21-cm power spectrum. This method has been explored in  \cite{2006ApJ...648..767M, 2009ApJ...695..183B, 2017NewA...57...94C} among others; it requires accurate knowledge of the foreground emissions.
Another approach is the foreground avoidance technique proposed and investigated by 
\cite{2010ApJ...724..526D, 2012ApJ...752..137M, 2012ApJ...745..176V, 2012ApJ...756..165P, 2013ApJ...776....6T, 2014PhRvD..90b3018L, 2014PhRvD..90b3019L, 2016ApJ...818..139T}, etc.
 It has been shown that, the foreground contamination remains within the ``Foreground Wedge'' \cite{2010ApJ...724..526D} if one estimates the  cylindrically averaged 21-cm power spectrum $P(k_{\perp}, k_{\parallel})$. Hence, in this technique  the cylindrical power spectrum  is measured in  the rest of the ``EoR-window'' in the $(k_{\perp}, k_{\parallel})$ plane where the foreground contamination is rather minimum.
%The Foreground Wedge corresponds to the cylindrical power spectrum $P(k_{\perp}, k_{\parallel})$ of the 21-cm brightness temperature fluctuations, where foregrounds are expected to dominate.
To enhance the accuracy of foreground removal, further foreground suppression methods are reported in literature \citep{2016MNRAS.463.4093C, 2019MNRAS.483.3910C, 2019MNRAS.483.5694B, 2021MNRAS.507.5310A}.  In these methods, the response of the primary beam is tapered using some window function which in turn suppresses the contribution from the bright sources lying in the outer regions of the telescope's field of view. Arora et al. (2021) \cite{2021MNRAS.507.5310A} discuss a gravitational-lense based power spectrum estimator that can suppress the Galactic diffuse synchrotron power spectrum by a few hundred times.

In addition to the significant challenge posed by foreground contamination, complex instrumental effects also play a crucial role in 21-cm observations. 
These instrumental effects can introduce various sources of systematic errors that need to be carefully addressed. Various techniques, almost always with a known sky model \citep{1984ARA&A..22...97P, 1992ExA.....2..203W,2007ITSP...55.4497V, 2009ITSP...57.3512W}, including primary calibration, self-calibration, and bandpass calibration, etc., 
are employed to estimate the gain and calibrate the observed visibilities.
%Given the array configuration and system temperature, in practiece, there is always a limited calibration accuracy. The residual gain and bandpass errors can significantly effect any high dynamics range observation.

The sky-based calibration techniques rely on  a good sky model, and the estimation of the gains is subjected to the sky model’s
accuracy, the telescope’s sensitivity, etc.
The unavailability of good calibrator sources in the sky, or imperfections in the sky model limits the accuracy of the gains which results in residual calibration/gain errors. The calibration accuracy is also limited due to various reasons such as antenna gain variations, instrumental electronics nonlinearities, spectral response variations, atmospheric effects at low frequency, RFI, etc \citep[e.g][]{2016MNRAS.461.3135B,  2016MNRAS.463.4317P}. The systematic errors in the data reduction process, and incomplete knowledge of instrumental effects, can also introduce errors if not performed correctly.
A few interferometers designed to have redundant baselines and a redundant baseline-based calibration approach is used \citep[e.g][]{2015ApJ...809...61A, 10.1093..mnras..staa3001}.

Numerous factors contribute to calibration errors that often restrict the detection of the redshifted 21-cm signal, and a mammoth effort has been invested to study and characterize the various effects of these errors.
In the context of LOFAR-EoR experiments, various calibration effects such as gain errors, polarized foreground contamination, ionospheric effects, and systematic biases arising from calibration processes have been thoroughly explored, and discussed by \cite{2018MNRAS.478.1484G, 2016MNRAS.463.4317P}. 

Additionally, studies by \citep{2015MNRAS.451.3709A, 2016JApA...37...35A, 2018MNRAS.476.3051A, 2015MNRAS.453..925V, 2016MNRAS.458.3099V,2016RaSc...51..927M} investigate the impact of polarization leakage and ionospheric effects.

In recent works Barry et al. (2016) \cite{2016MNRAS.461.3135B}, Ewall-Wice et al. (2017) \cite{2017MNRAS.470.1849E} and Byrne et al. (2019) \cite{2019ApJ...875...70B} have demonstrated that inaccuracies in sky-based calibration models, and issues related to non-redundancies within redundant calibration methods \citep{2018AJ....156..285J, 2019MNRAS.487..537O} can result in gain errors that can potentially contaminate the EoR window. Liu et al. (2010)\cite{2010MNRAS.408.1029L} demonstrated that non-redundancy in the baseline distribution leads to spectral artefacts that can contaminate EoR detection. Furthermore, Choudhuri et al. (2021) \cite{2021MNRAS.506.2066C} and Dillon et al. (2020) \cite{2020MNRAS.499.5840D} have investigated the effect of non-redundancy on gain solutions in redundant arrays, specifically the Hydrogen Epoch of Reionization Array (HERA). Kern et al.(2020) \cite{2020ApJ...890..122K} discuss calibration strategies for HERA and assess their impact on 21- cm power spectrum. They demonstrate that the unmodeled diffuse flux and instrumental contaminants corrupt the gain solutions, and discuss gain-smoothing procedures to mitigate these gain errors.
Instrumental effects can also manifest as beam-related distortions. The telescope's beam pattern can introduce spatial variations and side-lobe contamination and Choudhuri et al.(2018) \cite{2018AJ....156..285J}, Joseph et al. (2019) \cite{2019MNRAS.487..537O} , Orosz et al. (2021) \cite{ 2021MNRAS.506.2066C} and Chokshi et al. (2024) \cite{2024MNRAS.534.2475C} have examined issues like antenna position errors and variations in telescope beam pattern with time and frequency and their impact on calibration solutions.
The presence of non-redundant antenna beams in redundant baseline calibrations introduces chromatic errors in gain solutions. These errors can lead to the contamination of the EoR window with foreground power leakage, as described in studies by \citep{2022ApJ...941..207K, 2023ApJ...953..136K} for HERA.

The study presented in this paper is fourth in a series of works on the ``Calibration requirements for EoR observations''. In the first of our works \citep[henceforth Paper~I]{2020MNRAS.495.3683K}, we studied the effect of time-correlated residual gain errors through simulated observations for GMRT baseline configuration. We found that the residual gain errors introduce a bias in the power spectrum estimations for the visibility correlation based estimators. % that use the correlation of visibilities in nearby baselines.
 Considering contributions from various types of baseline pairs involved in the visibility measurement to be used in visibility correlation, we found that %for the visibility correlation-based power spectrum estimator,
the bias in the power spectrum arises mainly from those types of baseline pairs that have at least one antenna in common.
In the second work \citep[henceforth Paper~II]{2022MNRAS.512..186K} we developed an analytical framework to calculate the bias and uncertainty of the power spectrum in the presence of time-correlated residual gain errors, and system noise. 
Through simulated observations of uGMRT, we verified the analytical results.
 In \citep[henceforth Paper~III]{2024JCAP...05..068G} we first access the gain characteristics of our uGMRT Band~3 ($300 - 500$ MHz) observations towards the ELAIS-N1 field \citep{2019MNRAS.490..243C}. We found that residual time-dependent gain errors from most of the antennae follow Gaussian distribution and are not correlated across the different stokes or different antennae. We then use the gain characteristics and the foreground estimates along the same line of sight \citep{2019MNRAS.490..243C} to access the bias and uncertainity in \HI power spectrum detection. We also commented on the time estimates for \HI signal detection with the uGMRT given similar gain characteristics. 

In this work, we incorporate the residual bandpass gains in our analytical framework and estimate the bias and uncertainty in the Multifrequency Angular Power Spectrum (MAPS) in presence of strong foregrounds. Further, we use these analytical frameworks to understand various effects of residual time- and frequency-correlated gain errors. Rest of the  paper is organized as follows. In Section~\ref{sec:gainmodel}, we discuss the time- and frequency-dependence in  gains and present a model for the correlated residual gain and bandpass errors. In Section~\ref{sec:analyticbiasvar}, we first discuss estimation of MAPS through the Taper Gridded Estimator. We then discuss   various types of baseline pairs used for such estimation and  present analytical expressions of bias and excess uncertainty in  CLTGE. In section~\ref{sec:sky}, we discuss the foreground model and fiducial 21-cm signal used here. The telescope properties and observational aspects with SKA1-Low are discussed in section~\ref{sec:ska}. In section~\ref{sec:result}, we present the result.  We conclude with the discussions in section~\ref{sec:discussion}.

\section{Time and Frequency Dependent Gain Errors}
\label{sec:gainmodel}

The electric field of an electromagnetic wave emitted by a celestial source and incident on upper ionosphere of the Earth undergoes various transformations before it is detected. These transformations encompass the impact of the ionosphere, the electronic and geometric characteristics of antennas and receivers, as well as the behaviour of amplifiers employed in the signal chain. We refer to the combined effects of the ionosphere and the instrumental system's response to the incoming electric field from the sky as ``gain''. These gains are complex function of time and frequency  of the observation.
The measured visibility $\tilde{V}(\vec{U}_i, \nu)$, at an observation frequency $\nu$, for the $i^{th}$ baseline $U_i$ at a time $t$ can be given as \citep{1996A&AS..117..137H, SynthesisImaging}
\begin{equation}
  \tilde{V}(\vec{U}_i, \nu) = \tilde{\mathbb{G}}_i(t, \nu) \tilde{V}^S_i(\vec{U}_i, \nu) + \tilde{N}_i(\vec{U}_i)
  \label{eq:measurement}
\end{equation}
where $\tilde{\mathbb{G}}_i(t, \nu)$ is the time- and frequency-dependent baseline-based complex gain for the $i^{th}$ baseline; $\tilde{N}_i(\vec{U}_i)$ is the thermal noise and $\tilde{V}^S_i$ is the true-visibility. We use the term true-visibility to denote the visibility in the absence of any gain or noise. Note that the true-visibility is assumed not to change during the observation time here. This is due to the fact that the time-scale of change in the sky signal is much larger than the observation time \citep{1986isra.book.....T}. In this work, we assume that the signal we are interested in, is not polarized and the  instrument does not have any polarization leakage. The polarization leakage has a lesser effect than the direct gain errors, however, it may play an important role for very high dynamic range observations.  We plan to investigate it in a separate work. %(see Chapter \ref{Chapter3}).
In most practical cases we can write $\tilde{\mathbb{G}}_i(t, \nu)$ as
\begin{equation}
    \tilde{\mathbb{G}}_i(t, \nu) = \tilde{\mathrm{G}}_i(t) \tilde{\mathfrak{B}}_i(\nu).
    \label{eq:gaintnu}
\end{equation}
Here $\tilde{\mathrm{G}}_i(t)$ represents the time-dependence of baseline based complex gain, and $\tilde{\mathfrak{B}}_i(\nu)$ is the frequency-dependence of baseline based complex gain, for the $i^{th}$ baseline.

With the advent of software-based correlators, it is safe to assume that the primary source of the gains can be tagged to the antenna \citep{2017isra.book.....T}.
We can write both the time- and frequency-dependent part of the gain $\tilde{\mathrm{G}}_i(t)$ and $\tilde{\mathfrak{B}}_i(\nu)$, as arising from the individual gains of the pair of antennae A and B, in terms of antenna-based gains as
\begin{equation}
 \tilde{\mathrm{G}}_i(t)  = \langle \tilde{g}_A(t')  \tilde{g}^*_B(t')  \rangle \ ; \ \ \ \ 
    \tilde{\mathfrak{B}}_i(\nu) = \langle \tilde{b}_A(\nu')  \tilde{b}^{*}_B(\nu')  \rangle,
 \label{eq:gain}            
\end{equation}
where $\tilde{g}_A(t')$, $\tilde{g}_B(t')$ and $\tilde{b}_A(\nu')$, $\tilde{b}_B(\nu')$ are the time-dependent gain and the frequency-dependent gain of the individual antenna A and B. The frequency-dependency of individual antenna response is termed as the bandpass response. Here the functions $\tilde{g}_A(t')$ and $\tilde{b}_A(\nu')$, etc, are given in terms of the continuous time and frequency variables $t'$ and $\nu'$. The angle brackets in the gain and bandpass response represent the average over the integration time and channel width, respectively.  

In radio interferometric observations, it is reasonable to define the interferometric noise $\tilde{N}_i(\vec{U}_i)$ as a Gaussian random variable with zero mean. The noise is also uncorrelated across different baselines as well as frequency channels. The standard deviation of the real or imaginary part of the noise  in each visibility, $\sigma_N$, in terms of the source equivalent flux density (SEFD), frequency width of the channel ($\Delta \nu_c$), and integration time for each visibility ($T_{\rm int}$), can be expressed as follows \citep{1986isra.book.....T}
\begin{equation}
    \sigma_N = \frac{\text{SEFD}}{\sqrt{2\Delta \nu_c\,  T_{\rm int}}}.
    \label{eq:sigma_v}  
\end{equation}

In our previous works (Paper~I, II and III), we focused exclusively on the time-dependent component of the gain and extensively investigated the implications of time-correlated gain errors. In current study, we expand our analysis by incorporating the frequency-dependent bandpass response of the instrument, in addition to the time-dependent gains. We specifically examine the impact of both time- and frequency-correlated residual gain and bandpass errors.
In the following discussion, we provide a brief overview of the bandpass calibration and origin of the bandpass errors.

\subsection{Bandpass Calibration and Errors}
The bandpass response of the instrument refers to its sensitivity to different frequency components within a given observation bandwidth. Ideally, the instrument should have a flat and uniform response across the entire bandwidth of interest. However, in reality, imperfections and variations can lead to deviations from this ideal response. These deviations result in variations in the instrument's sensitivity at different frequencies, leading to the introduction of bandpass errors in the measured signals.
The frequency-dependent component of the gain, denoted as $\mathfrak{B}_i(\nu)$, is determined through the process of bandpass calibration. This calibration technique involves observing a bright source with a flat and known spectrum to correct for variations in amplitude and phase with frequency. Bandpass solutions are obtained for individual channels or by averaging a few nearby channels, depending on the signal-to-noise ratio of the observation.
Bandpass calibration serves multiple purposes. Firstly, it corrects for the frequency-dependent variations in the gains, ensuring a consistent amplitude and phase response across the observation bandwidth. Additionally, it helps in compensating for the slow variations of the bandpass response over time, thereby improving the overall calibration accuracy.

Accurate bandpass calibration is crucial for detecting and precisely measuring spectral features in the observed data. It plays a significant role in achieving a high dynamic range in continuum observations. However, there are several factors that can limit the accuracy of bandpass calibration, e.g.,  the availability of a suitable calibrator source with a well-known and stable spectrum during the observation. In cases where such a calibrator is not available, errors in the complex response of bandpass filters can introduce residual errors in the derived bandpass solutions.
Furthermore, the presence of radio frequency interference (RFI) or poor instrumental response in certain observation channels may require the removal of data from those channels. This can result in missing data for specific frequencies, which is often addressed by interpolating the bandpass solutions from neighbouring channels or using polynomial models to estimate the missing values. However, these interpolation techniques can introduce frequency correlations in the residual bandpass errors.
Further, such missing frequency channels introduce artefacts in the estimated power spectrum. In such cases, it is wise to estimate the  Multifrequency Angular Power Spectrum (MAPS) \cite{2007MNRAS.378..119D}, that depends on the angular multipole $\ell = 2 \ \pi U$ and the frequency separation directly from the visibilities. The MAPS-based Tapered Gridded Estimator (TGE) discussed by Bharadwaj et al. (2019)\cite{2019MNRAS.483.5694B},  Pal et al.(2021)\cite{2021MNRAS.501.3378P} presents a method to overcome this issue, where they first correlate the visibility data across frequency channels to estimate $C_{\ell}(\Delta \nu)$ at each frequency separation $\Delta \nu$ and then uses this $C_{\ell}(\Delta \nu)$ to estimate the power spectrum $P(k_{\perp}, k_{\parallel})$. In this work, we concentrate on estimating the MAPS and the bias and variance introduced into it due to residual gain and bandpass errors.

\subsection{Modeling Time- and Frequency-Dependent Gain Errors}
\label{subsec:modeling}
 In order to prepare the observed data for scientific analysis, it is necessary to estimate the antenna gains using calibration techniques.  In this work, we assume that optimal calibration methods, including primary calibration and self-calibration for time-dependent gain, as well as bandpass calibration to correct for frequency response, have been employed for the data under consideration. Once the calibrations are done, we assume that the antenna-based gain and bandpass terms $\tilde{g}_A$ and $\tilde{b}_A$  are unity, and so the gain term in eqn~\ref{eq:gaintnu} is also unity.  As the observations are limited by thermal noise, the time- and frequency-dependent gains can only be estimated to a limited accuracy. Furthermore, the finite time-cadence of the observation ($T_{\rm int}$) and the width of each frequency channel ($\Delta \nu_c$) only allow to limited estimation of the time- and frequency-correlation in gains. Hence, in reality, even after performing the calibrations, the terms $\tilde{g}_A$ and $\tilde{b}_A$ differ slightly from unity.  We call these as antenna-based ``residual gains'' and ``residual bandpass'' to distinguish them from the ideal unit gain cases after calibration. The residual gains and residual bandpass for each antenna contribute to the gain term $\tilde{G}_i(t)$ and $\tilde{\mathfrak{B}}_i(\nu)$, respectively.

Following a simplifying assumption in Paper~II, here we assume that the statistical properties of the residual gain errors as well as the residual bandpass errors are same for all the antennae. For an ideal interferometer with identical antennae/stations, the statistical properties of the gain from different antennae/stations are expected to be the same. In practice, though, the statistical properties of the residual gains from each antenna do differ. We here assume that we have allowed a minimum difference in the statistical properties of the antenna in the array and have not used any antenna for which the properties differ more than a threshold.  Henceforth, we drop the antenna suffix from $\tilde{g}_A(t)$ and $\tilde{b}_A(\nu)$ henceforth. For the time-dependent part of the gains $\tilde{g}(t)$, we use the  gain error model as discussed in Paper II. The gain from an individual antenna  is given as
\begin{equation}
 \tilde{g}(t) = \left [ 1+\delta_{R}(t) + i\delta_{I}(t)\right ]
 \label{eq:gaindef}
\end{equation}
where $\delta_{R}(t)$ and $\delta_{I}(t) $ are the real and imaginary part of the residual gain error from the antenna and they have  zero mean. We define 
\begin{equation}
\sigma^2_{\delta C} = \langle \delta^2_{C} \rangle, \ \ \ \ \eta_{C}(\tau) = \langle \delta_{C}(t) \delta_{C}(t+\tau) \rangle / \sigma_{C}^2,
 \label{eq:gainmod0}
\end{equation}
where $C$ is to be read as $R$ or $I$ for real and imaginary parts of the gain. Here the quantity  $\sigma^2_{\delta R}$ gives the variance of the real part of the residual time-dependent gain. Similarly, $\eta_{R}(\tau)$ gives the time-correlation in the residual gain. We further assume that the real and imaginary parts of the residual gain errors are not correlated, and the antenna gain properties from different antennae are also uncorrelated. We model the normalized two-point correlation $\eta(\tau)$ of the residual gains as
\begin{equation}
\eta(\tau) = \exp \left [ - \frac{\tau^2}{2 T_{corr}^2}\right ]
\label{eq:etadef}
\end{equation}
where $T_{corr}$ gives the correlation time of the residual gains. 

The frequency-dependent bandpass response for the individual antennae can be written as 
\begin{equation}
 \tilde{b}(\nu) = \left [ 1+b_{R}(\nu) + ib_{I}(\nu)\right ]
 \label{eq:banddef}
\end{equation}
where $b_{R}(t)$ and $b_{I}(t) $ stands for the real and imaginary part of the residual bandpass error. 

 Similar to the time-correlated gain errors, for the bandpass errors, we make the assumption that they are also Gaussian random variables with bandpass errors from all antennae following same statistics. The statistical properties of the bandpass errors are characterised by their variance and two-point correlation functions. Additionally, we assume that there is no correlation between the real and imaginary parts of the bandpass errors for a given antenna and the residual bandpass errors from different antennas are  uncorrelated. Note that the assumptions presented here for both time- and frequency-dependent response of the antenna need not be valid for an arbitrary array design. However, any presence of such cross correlation will incresae the systematics in detection of MAPS. Hence, in practice, we consider that for such observations we use well designed radio interferometers where the various cross-correlations are negligible. These assumptions are found to be true in our observation of uGMRT Band~3 in Paper~III.

We denote the variance of the bandpass $\sigma^2_{bC}$ and the normalized two-point correlation $ \xi(\Delta \nu)$ of the residual bandpass for a given antenna as
\begin{equation}
\sigma^2_{bC} = \langle b^2_{C} \rangle, \ \ \ \ \xi(\Delta \nu) = \langle b_{C}(\nu) b_{C}(\nu+\Delta \nu) \rangle / \sigma_{bC}^2,
 \label{eq:bandmod0}
\end{equation}
where subscript $b$ is for the bandpass. Here the quantity  $\sigma^2_{b R}$ gives the variance of the real part of the residual frequency-dependent gain and  $\xi_{R}(\Delta \nu)$ gives the frequency-correlation in the residual bandpass. Note that we have assumed that the normalised two-point correlation of the residual bandpass is a function of frequency separation $\Delta \nu$ only. We have also assumed that the normalised two-point correlation functions are the same irrespective of being of real or imaginary parts. We  model the residual bandpass error as
\begin{equation}
\xi(\Delta \nu) = \exp \left [ - \frac{\Delta \nu^2}{2 \nu_{\rm corr}^2}\right ]
\label{eq:xidef}
\end{equation}
where $\nu_{\rm corr}$ is the correlation frequency of the residual bandpass. 
In reality, the normalized two-point correlation function for the residual gain or bandpass errors can be rather complicated and needs to be estimated for a particular observation.

It has been shown in various works such as \cite{2016ApJ...818..139T, 2017MNRAS.470.1849E} that for spectrally smooth bandpass solutions, the power spectral bias from residual gain errors will not be significant. Due to the smoothness relative to the EoR spectral modes, the spectral contamination will primarily be confined to the wedge and the power leaked from the intrinsic foreground is not coupled to the EoR window \citep{2016MNRAS.461.3135B}. However, a departure from the spectrally smooth behaviour of the frequency-dependent gains may result in complex residual gain errors in frequency, and contaminate the  EoR window. Although the existing experiments enforce the spectral smoothness of the frequency-dependent gains, here we do not strictly make any such assumption for the development of the framework to estimate the residual effects. The framework that is being presented employs a simple Gaussian model for both time- and frequency-correlated gain errors, but it is not confined to this model and may be applied to any model, regardless of whether the gains are assumed to be spectral smooth or not. \footnote{We found that there exists residual gain correlation of $\sim 2 $ MHz in uGMRT band 2 and band 3 observations. These data are being investigated and will be presented separately.}

\section{Analytical Estimates of Bias and Variance of the redshifted 21- cm Power Spectrum}
\label{sec:analyticbiasvar}
In this section, we first discuss a TGE based estimator for MAPS and then  present a methodology to analytically estimate  the bias and variance in  estimating MAPS  for a known  time- and frequency-dependent gain and bandpass error model and a known model for the sky. 

\subsection{The Multifrequency Angular Power Spectrum Estimator}
Bharadwaj et al. (2001) \cite{2001JApA...22..293B}  shows that by correlating the visibilities at a given baseline and frequency channels we can estimate the 2-dimensional angular power spectrum (2DPS) of the \HI brightness temperature fluctuation in the sky plane at the redshift corresponding to the observing frequency. Using this idea, several variants of the visibility-based power spectrum estimators are implemented, e.g.  Bharadwaj et al. (2005) \cite{2005MNRAS.356.1519B},  Begum et al. (2006) \cite{2006MNRAS.372L..33B}, Choudhuri et al. (2014) \cite{2014MNRAS.445.4351C}, and others. These methods are used  to estimate the angular power spectrum of diffuse galactic foregrounds as well as the power spectrum of \HI in nearby galaxies \citep{2012MNRAS.426.3295G, 2017MNRAS.470L..11C, 2019MNRAS.490..243C, 2020MNRAS.494.1936C, 2009MNRAS.398..887D, 2013MNRAS.436L..49D,  2020MNRAS.496.1803N, 2023MNRAS.526.4690N}.  Choudhuri et al. (2016 a) \cite{2016MNRAS.459..151C} and Choudhuri et al. (2016 b) \cite{2016MNRAS.463.4093C} introduce a gridded estimator (TGE) that tapers the response of the telescope beam and hence suppresses effects of the point sources outside the primary beam that can not be imaged properly. In Paper~I and II, we  establish a methodology to assess the  bias and uncertainty in  2DPS based on the TGE and in Paper~III we implemented this to estimate the systematics  that arises through the time-dependent residual gain error in a uGMRT observation. The visibility based 2DPS is extended further in \cite{2007MNRAS.378..119D}, where they propose to correlate the visibilities from different frequency channels and hence find the frequency-correlation in the signal. This then measures the multifrequency angular power spectrum (MAPS) of the sky brightness temperature   as a function of both angular multipole ($\ell$) and frequency separation ($\Delta \nu$). The sky brightness temperature distribution $\delta T(\hat{n}, \nu)$ is first decomposed into its'  spherical harmonics $a_{lm}(\nu)$, the  MAPS $C_{\ell}(\Delta \nu) $ is defined as
\begin{equation}
  C_{\ell}(\Delta \nu) = \langle a_{lm}(\nu) a_{lm}^*(\nu+\Delta \nu) \rangle,
  \label{eq:cldef}
\end{equation}
where $\Delta \nu$ is the difference in frequency channels for correlation. Note that the angular multipole $\ell$ is related to the baseline $U$ as $\ell = 2 \pi U$, and hence, each baseline measure one angular multipole. Here we assume that the sky brightness temperature fluctuations are statistically homogeneous and isotropic  and  depends only on the frequency separation $\Delta \nu$ between the two frequency channels for correlation. This assumption of homogeneity in frequency, helps us to  average  estimates of MAPS from different frequency-pairs  giving the same frequency separation. During the epoch-of-reionization, however, the mean neutral hydrogen density changes rapidly and hence the MAPS should be considered as a function of the frequency pairs at which the correlations are performed \citep{2019MNRAS.483L.109M}. Estimating the non-homogenous part of the signal, requires to measure it with much higher accuracy and is rather more challenging \citep{2020MNRAS.494.4043M}. Here, we restrict ourselves to the estimator of MAPS that assumes the signal to be ergodic. 

Bharadwaj et al. (2019) \cite{2019MNRAS.483.5694B} develops a TGE based estimator, that can use measurement of  visibilities in different frequency channels to estimate the MAPS assuming homogeneity of the signal in frequency, we will call this as CLTGE henceforth. This estimator is then used to estimate the spherically or cylindrically averaged power spectrum. Here we outline the basic working of CLTGE. The visibilities are first gridded in baselines for all the  frequency channels. The grid size $\Delta U$ is chosen such that $\Delta U < (\pi \theta_0)^{-1}$, for a telescope with field of view extended to $\theta_0$ radians. The visibility correlations are then calculated within each grid correlating visibilities from different channels. The gridding process also implements a convolution kernel that tapers the primary beam response to avoid effects from foregrounds outside the field of view of observation. For the zero-frequency-separation $\Delta \nu = 0$,  the visibility correlation is performed  only with different baselines in a given grid to avoid noise bias. The correlations with same frequency separations $\Delta \nu$ for each baseline-grid are then identified and their average is estimated. These are then further averaged over a given annular region in the baseline plane to get the estimates for $C_{\ell}(\Delta \nu)$. The averaging process ensures that proper weights are used to incorporate the effect of incomplete baseline coverage as well as the tapering kernel. It has been shown \citep{2019MNRAS.483.5694B}  that in absence of residual gain and bandpass errors, CLTGE gives an unbiased estimate. In this work, we  access the bias and excess uncertainty  introduced by time-dependent residual gain errors and frequency-dependent bandpass errors in CLTGE. %,  

\subsection{Baseline Pair Fractions}
\label{subsec:newBPF}
In interferometry, a baseline vector is defined as the antenna separation in units of the observed wavelengths. The sky-plane is considered as the tangent plane passing through the centre of the field of view in the sky. The baseline-plane where the visibilities are gridded is on the earth's surface and   is parallel to the sky plane. Here, we consider that all the antennae of the interferometer array are tracking the source.  As the source position changes with time in the sky, the baseline vector assumes different components in the baseline plane. Hence, a pair of antennae $A$ and $B$ provides measures of visibility at different baselines at different times during the observation. We denote  the visibility measured by a pair of antennae $A, B$ at a time $t$  and frequency $\nu$ as $\tilde{V}_{AB}(t, \nu)$.  The work presented in Paper I, II and III can be considered as estimation of bias and uncertainty in $C_{\ell}(\Delta \nu=0)$. Observationally, correlating the visibilities from the same frequency channel shows a sudden jump in the power \citep{2021MNRAS.501.3378P}. This is because bandpass responses in visibility correlations from same frequency channels can be  self-correlated. We do not consider the correlation of the visibilities measured at the same frequency channel here. The visibility correlation in a given grid can be obtained with different types of antenna-pairs and time, giving rise to types of baseline-pairs. In absence of residual gain and bandpass errors, all the baseline pairs in a grid can be considered equivalent, irrespective of the antenna pairs and time they originate from. However, baseline pairs of different types have slightly different contributions, if the residual gain and bandpass errors are considered. We define baseline-pair-fraction as  the ratio of   a given type of  baseline pairs to the total number of baseline pairs in a given grid and denote it as $n_{i}'$ for different types $_{i}$. We list all the different possible baseline pairs here:
\begin{itemize}
    \item {\bf Type 1} Correlation of the visibilities measured by the same antenna pair, at different times, and different frequencies, i.e.
      $\langle  \tilde{V}_{AB}(t, \nu) \tilde{V}^*_{AB}(t', \nu')\rangle$. The fraction of such baseline pairs is $n_{1}'$.
     \item {\bf Type 1A} Correlation of the visibilities measured by the same antenna pair, at the same time, and different frequency, i.e.
      $\langle  \tilde{V}_{AB}(t, \nu) \tilde{V}^*_{AB}(t, \nu')\rangle$. The fraction of such baseline pairs is $n_{1A}'$. 
    \item {\bf Type 2} Correlation of the visibilities measured by antenna pairs having one antenna in common, measured at the same time, and different frequency, i.e.
    $\langle  \tilde{V}_{AB}(t, \nu) \tilde{V}^*_{AC}(t, \nu')\rangle$. The fraction of such baseline pairs is $n_{2}'$.
    \item {\bf Type 3} Correlation of the visibilities measured by antenna pairs having one antenna in common measured at different times, and different frequency $\langle  \tilde{V}_{AB}(t, \nu) \tilde{V}^*_{AC}(t', \nu')\rangle$. The fraction of such baseline pairs is $n_{3}'$.
    \item {\bf Type 4} Correlation of the visibilities measured by antenna pairs having no antenna in common measured at any time.
     $\langle  \tilde{V}_{AB}(t, \nu) \tilde{V}^*_{CD}(t', \nu')\rangle$. The fraction of such baseline pairs is $n_{4}'$.
\end{itemize}
Note that, the noise is uncorrelated for any two measurements taken into account for the aforementioned cases.  

\subsection{Bias and Variance of the Angular Power Spectrum}
\label{subsec:BVAPS}
Several methods have been developed to reduce  foreground contamination from EoR observations. One of the most commonly used techniques is called ``foreground subtraction'' \citep[e.g.][]{2006ApJ...648..767M, 2009ApJ...695..183B, 2017MNRAS.470L..11C}, which involves estimating the foreground and then subtracting them from the total observed signal. In such methods, the visibilities from the foreground sources are estimated and then subtracted from the observed visibilities. On the other hand, for the extended foreground emissions, MAPS for the foreground is estimated and then subtracted. 
Another approach is to use ``foreground avoidance'' techniques \citep[e.g.][]{2010ApJ...724..526D, 2012ApJ...752..137M, 2012ApJ...745..176V}, which involve designing observational strategies that avoid frequency ranges where the foreground signals are strongest. This approach is based on the fact that the foreground signals have a different spectral signature than the EoR signal, and hence can be avoided by restricting to places in the $(k_{||}, k_{\perp})$ plane.
To enhance the accuracy of foreground removal, further foreground suppression methods are reported in literature \citep[e.g.][]{2016MNRAS.463.4093C, 2019MNRAS.483.3910C, 2019MNRAS.483.5694B, 2021MNRAS.507.5310A}. In this method, we taper the response of the telescope beam using  a Gaussian kernel to  suppress the contribution from the bright sources lying in the outer regions of the telescope's field of view. 
Here, we assume that we already have an adequate estimate of the foreground and a foreground mitigation method is applied to the data. We do not include any error that may arise through the estimation of the foreground itself.

The estimator CLTGE gives unbiased estimates of MAPS in absence of residual gain and bandpass errors \citep{2016MNRAS.463.4093C}. Uncertainties in  CLTGE are estimated by simulating various realizations of the  system noise and sky brightness temperature fluctuations. The uncertainty in CLTGE $\sigma_{\rm T}$ includes the effect from  the sample and cosmic variance and the contribution from the system temperature. In absence of gain errors, for a  21-cm MAPS $C_{\ell_{HI}} (\Delta \nu)$,  $\sigma_{\rm T}$ can be approximately given as
\begin{equation}
  \sigma^2_{\rm T}  =  \frac{C^2_{\ell_{HI}}}{N_G} + \frac{N_2\ C_{\ell_{HI}} }{N_B N_d} + \frac{N_2^2}{2N_B N_d^2},
   \label{eq:sigmaT} 
\end{equation}
which we denote as thermal part of the uncertainty henceforth. The quantities $\sigma_{\rm T}, N_G, N_B$ are function of $\ell$, the $\Delta \nu$ dependence on $\sigma_{\rm T}$ arises from the same in $C_{\ell_{HI}}$. Note that various quantities given in this expression can vary with both the angular multipole $\ell$ and $\Delta \nu$. Here, we assume that for each day, the observations are made for $N_h$ hours and all over $N_d$ days of  observation the same baselines are repeated. Hence, the total observation hours would be $T_{obs} = N_h \times N_d$. The variables $N_G$ and $N_B$ represent the count of independent estimates of MAPS in an annulus and the total number of visibility correlations within a baseline annulus in a given day. The quantity $N_2$ encompasses the effect of system temperature and will be discussed shortly.

Presence of emission from  sources, other than the cosmological \HI, in the same observing frequency  are collectively called foregrounds. As these are several orders of magnitude higher than the expected redshifted 21-cm signal, measuring the later turns to  a problem of high dynamic range interferometric observation. Since emissions from various type of sources are not correlated, the measured MAPS can be considered as a summation of the MAPS due to the redshifted 21-cm signal and that from the foreground.

Presence of foregrounds in the observed visibilities introduces a second order effect.  The residual gain and bandpass errors, in presence of various foreground emissions, enhance the  uncertainty in CLTGE  and introduce a non-zero bias. Once the foreground mitigation methods are applied to the observed MAPS, this second order effect becomes important.  Paper II discuss the analytical framework that uses several  assumptions, to estimate the bias and uncertainty in MAPS with $\Delta \nu =0$. As the present work uses the same assumptions and methodology, we do not repeat it here,  and  refer the reader to section~2.4 of Paper II  for details.   The bias and uncertainties in CLTGE as calculated for the residual gain and bandpass error models described in section 2.2 of this paper is given below. Here we denote the bias as $\mathcal{B}_{C_{\ell}}$ :
\begin{equation}
\mathcal{B}_{C_{\ell}}  =  \left[ (n_{13}\chi + n_{12}) \Sigma_2^{\delta+} + (n_{13} + n_{12}) \Sigma_2^{b+} \xi \right]  \frac{C_{\ell}}{N_d}
\label{eq:bias}
\end{equation}
where $C_{\ell}(\Delta \nu)$ is the combined angular power spectrum of the foreground ($C_{\ell_{FG}}$)  and redshifted 21-cm signal ($C_{\ell_{HI}}$). However, since the latter is much weaker, we can safely neglect its effect. Note that, in  eqn~\ref{eq:bias}, we have not explicitly show the dependence of $\ell$ and $\Delta \nu$ on various quantities. Here $n_{13} = 2n_1' + n_3' $ and  $n_{12} = 2n_{1A}' + n_2' $, both depends on the angular multipole $\ell$, $\Sigma_2^{\delta \pm} = \sigma^2_{\delta R} \pm  \sigma^2_{\delta I}, \Sigma_2^{b \pm} = \sigma^2_{b R} \pm \sigma^2_{b I}$ are independent of $\ell$. The quantity $\xi$ depends on $\Delta \nu$ as in eqn~\ref{eq:xidef}.
The variable $\chi$ encompasses the effect of  time-correlation of the residual gain error and is given as
\begin{equation}
  \chi(\ell) = \frac{1}{T^2_D}\int^{T_D}_{T_{\rm int}} \ \left [ T_D - \tau \right ] \eta(\tau) \ {\rm d} \tau,
   \label{eq:chidef} 
\end{equation}
with $T_D = \frac{\Delta U T_{24}}{\ell}$. Here $T_{24}$ corresponds to one sidereal day. It is clear that the bias is directly proportional to the MAPS being measured. The bias reduces with number of  observation  days. We have assumed that the residual gain errors do not have any long term time-correlation. Note, such correlations can arise because of gravity loading in antenna dishes or imperfect beam-forming and need to be investigated at the telescope facilities. In such cases, the bias and $\sigma_E$ would not scale down with $N_d$ as shown here. Apart from the multiplicative contribution from the foreground MAPS, the bias depends on the angular multipole $\ell$ through the  baseline pair fractions and the function $\chi$. The $\Delta \nu$ dependence in the first term of the bias $\mathcal{B}_{C_{\ell 1}} = (n_{13}\chi + n_{12}) \Sigma_2^{\delta+}  \frac{C_{\ell}}{N_d}$ comes through the frequency dependence of the foreground MAPS.  The additional $\Delta \nu$ dependence on bias arises from the second term $\mathcal{B}_{C_{\ell 2}} = (n_{13} + n_{12}) \Sigma_2^{b+} \xi  \frac{C_{\ell}}{N_d}$.
Elahi et al. (2023)\cite{2023MNRAS.525.3439E} implement foreground suppression in CLTGE using the fact that the frequency de-correlation  is much slower in the foreground than the 21-cm signal. They estimate the foreground characteristics at $\Delta \nu$ values at which the redshifted 21-cm signal is expected to have no correlation and use it to subtract the foreground in the MAPS at lower $\Delta \nu$. Here, the $\Delta \nu$ dependence in the first term in the bias expression comes directly from the $\Delta \nu$ dependence in the foreground contribution to measured MAPS. Hence, this foreground suppression method result in suppression of the first term in bias. 

We denote the  uncertainty in CLTGE as $ \sigma_{C_{\ell}}$, where the excess uncertainty due to the residual gain and bandpass errors are denoted as  $ \sigma_{E}$, with  $ \sigma^2 _{C_{\ell}} = \sigma^2_{T} + \sigma^2_{E}$. The excess uncertainty then can be written as 
\begin{eqnarray}
 \sigma^2_{E} &=&  2\  \frac{\mathcal{B}_{C_{\ell}}^2}{N_G} + 2 \left [  \Sigma_2^{\delta + } + \Sigma_2^{b+} \right ] \frac{N_2 C_{\ell}}{N_B N_d^2} \\ \nonumber
&+& 4 \left [  (\Sigma_2^{\delta + } + \Sigma_2^{b +} )^2 +  (\Sigma_2^{\delta - } + \Sigma_2^{b -} )^2 \right ]  \frac{C_{\ell}^2}{N_G N_d^2} \\ \nonumber
 &+& \left[ (n_{13}\chi + n_{12}) \Sigma_2^{\delta-} + (n_{13} + n_{12}) \Sigma_2^{b-} \xi \right]^2  \frac{C_{\ell}^2}{N_G N_d^2}.
  \label{eq:sigmaE} 
\end{eqnarray}
As discussed earlier, for all practial purposes $C_{\ell} \sim C_{\ell_{FG}} $. Note that, if the values of $\sigma_{\delta R} = \sigma_{\delta I}$ and $\sigma_{b R} = \sigma_{b I}$, $\Sigma_2^{\delta - } =  \Sigma_2^{b -} = 0$, the expression for variance above becomes simpler. We have also assumed that all the quantities that contribute to the residual gain and bandpass errors have exactly same statistical property for both the polarisations $RR$ and $LL$ usually used for such observations. The quantity  $N_2$ is the noise correlation between visibility pairs. Since the noise in different visibilities is uncorrelated, in case of CLTGE with $\Delta \nu = 0$, $N_2  = \langle \tilde{N_i} \tilde{N_j}^{*} \rangle = {2\sigma_N^2 \delta _{ij}}$. The Kronecker Delta $\delta_{ij}$ ensures that the noise-correlation between any two antennae $i, j$ is zero and the correlation is $N_2$ when the same antenna-pair is used.
When correlating the visibilities across different channels, it is essential to consider that the number of independent samples of noise correlation will vary depending  on the frequency-separation  $\Delta \nu$. Specifically, when nearby channels are considered, a greater number of channel pairs are available, resulting in a higher count of independent samples of noise correlations across the bandwidth (BW) of the observation. However, as the frequency-separations increase, the number of independent samples of noise correlations decreases. For instance, when $\Delta \nu \sim {\rm BW}$, only one channel pair remains available for such averaging.
Hence, it is important to note that the noise correlation exhibits a dependence on $\Delta \nu$, and its behaviour can be characterized as a function of this parameter as:
\begin{equation}
    N_2(\Delta \nu) = \frac{\sigma_{N}^2}{(BW- \Delta \nu)/\Delta \nu_c},
    \label{eq:noisecorr}
\end{equation}
where $\Delta \nu_c$ is the width of each frequency channel.

\section{Modeling the sky signal}
\label{sec:sky}

In this work we aim to access the effect of residual gain and bandpass errors in observing the redshifted 21-cm signal from the Epoch of Reionization using the upcoming telescope SKA1-Low. Here we choose a fiducial redshift of $z_0 = 8$ that corresponds to an observing frequency of $\nu_0 = 157.8$ MHz and a bandwidth of ${\rm BW} = 50$ MHz. In this section, we discuss the foreground and fiducial 21-cm signal model that is used for the estimation and further analysis of bias and uncertainty in CLTGE.
\subsection{Foreground Model}
\begin{figure*}
    \centering
    \includegraphics[width=0.98\textwidth]{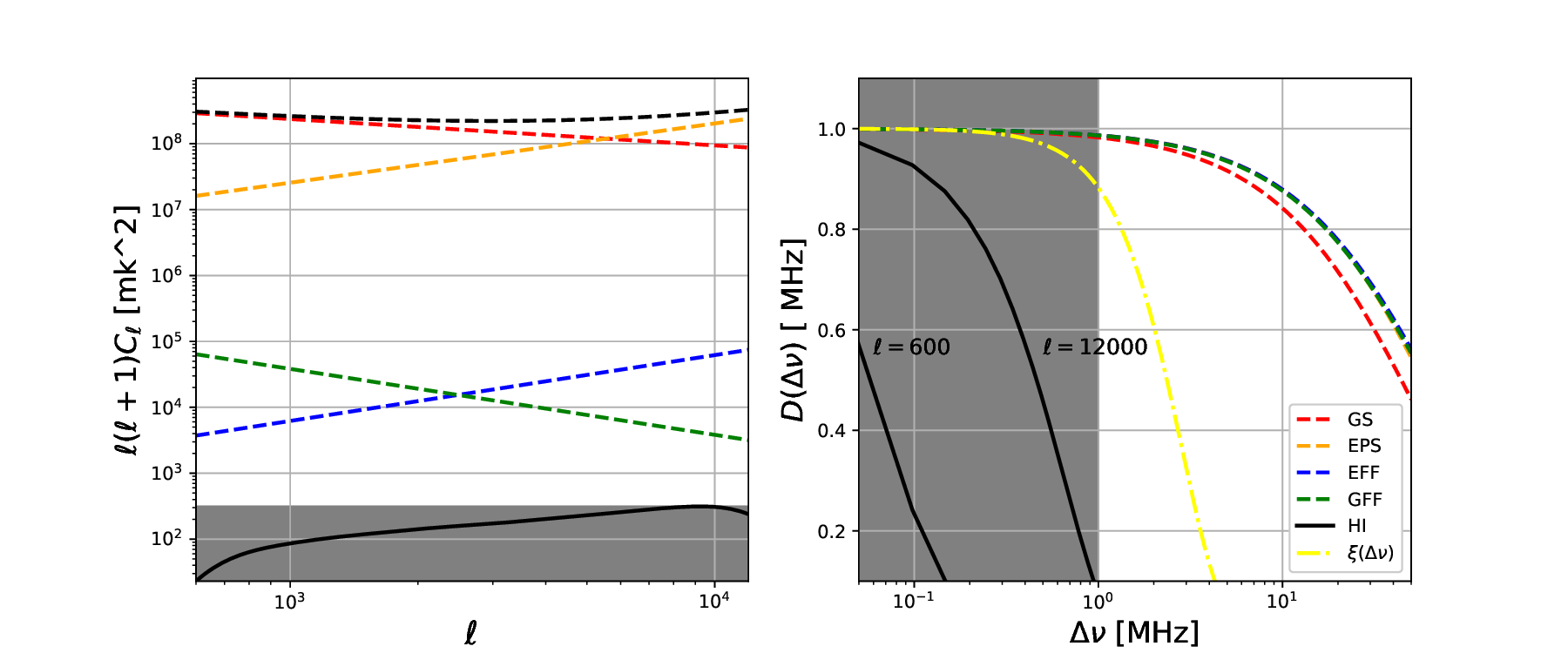}
    \caption{Variation of MAPS $C_{\ell}(\nu, \Delta \nu)$ as a function of angular multipole $\ell$ and frequency separation $\Delta \nu$  for models of foreground and redshifted 21-cm signal. Left panel plots the $\ell ( \ell + 1 ) C_{\ell}(\nu, 0) $ as a function of $\ell$ at a frequency of $\nu = 157.8$ MHz, which  corresponds to a  redshift of $8$ for the 21-cm signal. The dashed lines show the contribution from different components of the foregrounds (see the legend in the right hand panel), the black solid line is the expected 21-cm signal at this redshift. The black dashed line show the total contribution by the foreground. In the right panel, we show the frequency-de-correlation $\mathcal{D}(\nu, \Delta \nu)$ of various components of the foreground (see the legends)  and the 21-cm signal at the same frequency $\nu$. Frequency-de-correlation of the foreground is independent of the angular multipole, however, for the redshifted 21-cm signal, the frequency-de-correlation depends on $\ell$. We have shown the de-correlation function for two $\ell$ values of $600$ and $12000$ for the 21-cm signal. The dot-dashed line show the function $\xi(\Delta \nu)$ for $\nu_{\rm corr} = 2$ MHz.}
    \label{fig:Cl_FG_HI}
\end{figure*}

\begin{table}
\begin{center}
\begin{tabular}{l|c|c|c|c}
\hline 
&$A ({\rm mK}^2)$ & $\beta$ & $ \alpha$  & $\gamma$ \\
\hline 
Galactic synchrotron          &  700  & 2.4 &  2.80 & 4.0 \\ 
Extragalactic point sources   &  57.0  & 1.1 &  2.07 & 1.0 \\ 
Galactic free-free          &  0.088  & 3.0 &  2.15 & 35 \\ 
Extragalactic free-free        &  0.014  & 1.0 &  2.10 & 35 \\ 
\hline
\end{tabular}
\end{center}
\caption{The values of the foreground parameters in eqn~(\ref{eq:fg}) for $\nu_f = 130$ MHz. The parameters are  reproduced from \cite{2005ApJ...625..575S}.}
\label{tab:fgl}
\end{table}
Various physical processes contribute to the foreground signal, these are Galactic synchrotron (GS), Extragalactic point sources (EPS), Extragalactic free-free (EFF) and Galactic free-free(GFF). A detailed discussion on these are available in  \citep{ 2002ApJ...564..576D, 2008MNRAS.385.2166A, 2008MNRAS.389.1319J, 2012MNRAS.426.3295G, 2010MNRAS.409.1647J, 2017MNRAS.470L..11C, 2021MNRAS.500.2264H}, etc.
For the analysis presented in this paper, we use  the foreground model presented in \cite{2005ApJ...625..575S} and we discuss it briefly.  
The expression for $C_{\ell_{FG}} (\nu, \Delta \nu)$ can be written as
\begin{equation}
C_{\ell_{FG}} ( \nu, \Delta \nu) = C_{\ell_{FG}} ( \nu, 0) \ D_{FG} ( \nu, \Delta \nu),
 \label{eq:clfg} 
\end{equation}
where $C_{\ell_{FG}} ( \nu, 0)$ gives the foreground MAPS at zero frequency separation and $D_{FG} ( \nu, \Delta \nu)$ gives the frequency-de-correlation in the signal. The former provides the $\ell$ dependence whereas the latter gives $\Delta \nu$ dependence.  For the $i$-th component of the foreground, these can be modelled as
\begin{eqnarray}
C_{\ell_{FG}}(\nu, 0) &=& A_i \left ( \frac{1000}{\ell}\right)^{\beta_i} \left(\frac {\nu_f}{\nu}\right)^{2 \alpha_i}  \\ \nonumber
D_{FG} ( \nu, \Delta \nu) &=& \exp \left ( - \frac{ \left [ \log ( 1 + \frac{\Delta \nu}{\nu} ) \right ] ^2} {2 \gamma_i ^2} \right ) \ \left (  1 + \frac{\Delta \nu}{\nu} \right )^ {-\alpha_i},
\label{eq:fg}
\end{eqnarray}
where $\nu_f = 130$ MHz is a reference frequency. The values for the parameters $A_i, \alpha_i, \beta_i, \gamma_i$ are given in Table~\ref{tab:fgl}. 

Contributions from different foreground components are shown in Figure~\ref{fig:Cl_FG_HI}. The left panel show variation of $\ell(\ell+1)C_{\ell_{FG}}(\nu, 0)$ as a function $\ell$. Clearly, the most dominant component at lower $\ell$ is the contribution from GS, whereas at larger $\ell$ the EPS component dominates. The contributions from EFF and GFF are rather small. In this work, we  combine the effect of all these components and use that as a foreground model. The right panel of Figure~\ref{fig:Cl_FG_HI} show the frequency-de-correlation factor $D_{FG} ( \nu, \Delta \nu)$. We note that the for various components of the foreground, the frequency-de-correlation is almost similar. Henceforth, unless it is required, we would not write the $\nu$ dependence of $C_{\ell}$ or $D$ explicitly.

\subsection{Redshifted 21-cm signal}

Section~3 of \cite{2017MNRAS.464.2992M} have outlined a model for the redshifted 21-cm power spectrum at the EoR, we briefly discuss it here.
The EoR 21-cm signal is generated in three steps: First,
a particle-mesh N-body code is used to simulate the dark matter distribution
in a comoving volume of $[215 {\rm Mpc}]^3$ with $3072^3$ grids to the observed redshift (in our case $z_o = 8$). From this simulated dark matter distribution at $z_o = 8$, they identify haloes with a minimum mass of $M_{min} = 1.09 \times 10^9 M_{\odot}$  using the Friends-of-Friends (FoF) algorithm \cite{1985ApJ...292..371D}. Finally, they use the excursion set formalism \citep{2004ApJ...613....1F} to generate ionization maps, assuming that ionizing sources are hosted in those halos and hydrogen traces dark matter. They map the redshift space 21-cm brightness temperature map  using peculiar velocities. In this model, the ionization of the neutral hydrogen is controlled by two key parameters, $N_{\rm ion}$ the ionizing photon production efficiency, and $R_{\rm mfp}$ the mean free path of ionizing photons. These parameters are set as $N_{\rm ion} = 23.21$ and $R_{\rm mfp} =20$ Mpc  to achieve $50$ \% ionization at $z_o=8$, which yields a Thomson scattering optical depth $\tau = 0.057$ (consistent with \cite{2016A&A...596A.108P}), and the reionization in this model ends at $z\sim6$   (see \cite{2001AJ....122.2850B}). They use fifty independent realizations of the simulation  to create a statistically independent signal ensemble, and these were used to estimate the 21-cm power spectrum $P_{HI}(k)$. 

Considering the spherically averaged power spectrum model $P_{HI}(k)$, we generate the corresponding cylindrically averaged power spectrum $P_{HI}(k_{\perp}, k_{||})$, where $k_{||}$ denotes the component of wave vector parallel to the line of sight of observation and $k_{\perp}$ corresponds to the component in the plane of the sky.
In this work, we neglect various redshift dependent effects in the observed \HI power spectrum including the redshift space distortion \citep{2013MNRAS.434.1978M, 2018MNRAS.476...96S} and the light cone effects \citep{2011MNRAS.413.1409M}.  We further calculate the fiducial \HI MAPS using
\begin{equation}
C_{\ell_{HI}}(\Delta \nu) = \frac{1}{\pi r_c^2} \int_0^{\infty} {\rm d} k_{||}\ \cos \left ( k_{||} r_c' \Delta \nu\right ) \ P_{HI}(k_{\perp}, k_{||}),
 \label{eq:clHI} 
\end{equation}
where $r_c$ is the comoving distance to the redshift $z_0$ and $r_c' = \frac{{\rm d} r_c}{{\rm d} \nu}$. 

The angular multipole $\ell$ dependence of the fiducial redshifted 21-cm signal from a redshift of $8$ is shown in the left panel of   Figure~\ref{fig:Cl_FG_HI}. Clearly, the signal is much weaker than the foreground as expected. The 21-cm signal de-correlation, however, depends on the angular multipoles and is shown for two angular multipoles $\ell=600$ and $\ell=12000$ in the right hand side of Figure~\ref{fig:Cl_FG_HI}. The signal is expected to de-correlate faster for higher values of $\ell$.

\section{Telescope properties: SKA1-Low}
\label{sec:ska}
The SKA1-Low, located at the Murchison Radio-astronomy Observatory (MRO) site in Western Australia, is an aperture phased array that is presently under construction \citep{2013ExA....36..235M, 2015aska.confE...1K}. The telescope is expected to start its early shared risk  science run (configuration AA4) in 2029. Once completed, the SKA1-Low is designed to have 512 stations, with the largest baseline spanning approximately 65 km. Each of these stations will have   256 log-periodic dual-polarized antennas giving an effective field of view of  $327^{'}$ at $110$ MHz.    The available bandwidth of operation will be $50- 350$ MHz with a uniform resolution channel width of $5.4$ kHz. Though the high frequency-resolution is essential for Radio Frequency Interference detection and mitigation, the foreground mitigation, as well as 21-cm signal detection inherently do not require such high frequency-resolution. In this work, we choose  $1024$ channels giving rise to a native channel separation of $\Delta \nu_c = 48.82$ kHz.   Given the highest baseline, this is well bellow the limit expected  bandwidth smearing limit. Considering time-width-smearing, the $T_{\rm int}$ for the SKA1-Low is to be kept below $2$ sec, we choose an integration time of $T_{\rm int} = 1$ sec for our analysis. This keeps the time-smearing low, as well not enhance the noise in each visibility. The exposure time calculator \footnote{\url{https:://sensitivity-calculator.skao.int/low}} gives an expected value of $\sigma_V = 63.7$ mK / channel. As mentioned earlier, the baseline grids used for CLTGE $\Delta U < ( \pi \theta_0)^{-1}$, given that $( \pi \theta_0)^{-1} \sim 5\ \lambda$, we choose $\Delta U = 4\ \lambda$. In this work, we have assumed the SKA antennae are
identical circular apertures with the above given specifications. We do not
use any beam simulations that consider the chromaticity,
asymmetry or rotation of the beam. These effects, however, are effectively
modelled in the time- and frequency-dependent correlated residual gain errors. 

Baseline pair fractions as discussed in section  3.2 depends on the antenna configurations, $T_{int}$, $\Delta U$, as well as the phase centre of observation in the sky. For this work, we assume that the observation pointing is at a right ascension of [$13h\ 31m\ 08s$]  and declination of [$-50^{\circ}\ 00'\ 00''$]. Using these, we generate the SKA1-Low baseline configuration for $N_h = 8$ hours of observation. We  then grid the baselines with a grid-size of $\Delta U$ as mentioned above to estimate the baseline pair fractions in each grid. Alongside we also estimate the total number of baseline $N_B$ in each grid. We have observed in Paper I that the \HI signal falls rapidly beyond a baseline of $1$ k$\lambda$, that is $\ell \sim 6000$ at $150$ MHz. We restrict our range of  $\ell$ between $600-12000$ for this work. The lower limit of this range is chosen to accommodate enough number of estimates of the CLTGE in the lowest bin to avoid the effect of sample variance. The limit at the higher end is taken to be twice the value of the angular multipole at which the \HI signal drops significantly. We further generate annular bins in $\ell$ range and estimate the average values of the baseline pair fractions, $N_B$ and the number of independent estimates of MAPS $N_G$ in each annular bin.

\begin{figure}
    \centering
    \includegraphics[width=0.9\textwidth]{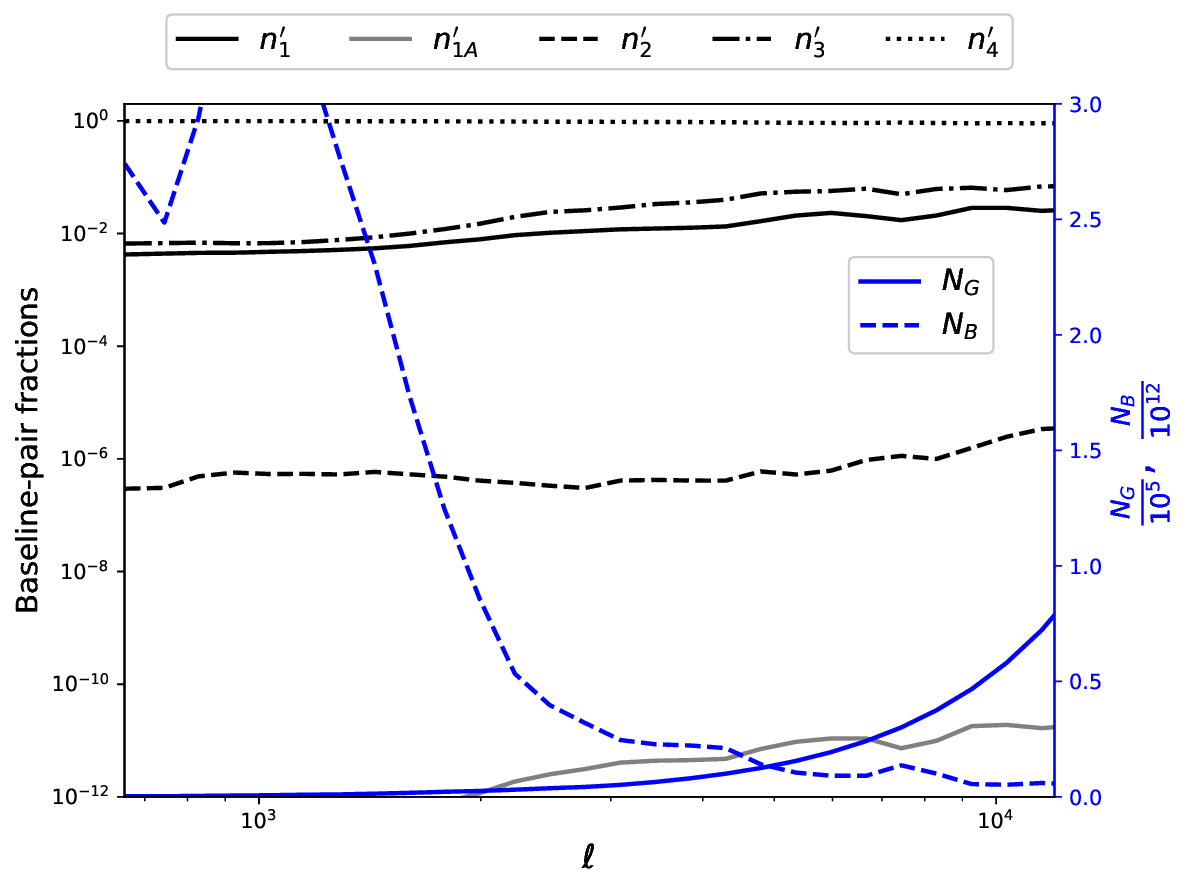}
    \caption{Variation of baseline pair fractions $n'_{i}$ with angular multipole $\ell = 2 \pi U$ for 8 hours of observation with the SKA1-Low. Integration time for the observation is kept at $1$ sec and uv-grid size is $0.004$ $k\lambda$. The fraction of baseline pairs is dominated by Type 4 for the above-used parameters. Number of grid points $N_{G}$ and average number of baselines in a given grid is also plotted as a function of $\ell$. Both these values are scaled, see the right-side axis (blue) for details. }
    \label{fig:ska_BPF}
\end{figure}

In  Paper~I, II we discuss the baseline pair fractions for visibility correlation within same frequency channels. The corresponding baseline pair fractions are similar to that in the Types 1, 2, 3 and 4 here, however, in the present case, the correlations between visibilities are across different frequency channels. The baseline pair fractions in Paper~I and II will be mentioned as  $n_{i}$ (that is without a prime). Here, we additionally have the baseline pair fraction Type 1A, where  visibilities are measured by the same antenna pair at the same time but different frequencies are considered. Assuming all the channels in a baseline-grid have the same number of baselines, with number of baseline in a given grid in a given channel as  $N_B$, the baseline pair fractions in Paper~I and II can be related to the baseline pair fractions defined here:
\begin{align}
    n_1' = n_1 \frac{N_B-1}{N_B} \ \sim n_1
    \\
    \nonumber
    n_{1A}' = \frac{1}{N_B} 
    \\
    \nonumber
    n_2' = n_2 \frac{N_B-1}{N_b} \ \sim n_2
    \\
    \nonumber
    n_3' = n_3 \frac{N_B-1}{N_B} \ \sim n_3
    \\
    \nonumber
    n_4' = n_4 \frac{N_B-1}{N_B} \ \sim n_4.
     \label{eq:nidef} 
\end{align}
The five different types of baseline-pair fractions are plotted in Figure~\ref{fig:ska_BPF}. In Paper~III we presented estimation of bias and  uncertainty for  the Band~3 uGMRT observation, where we found that baseline-pairs of Type 1 and 3 contribute more than 50\% in all the grids at different $\ell$. We observe that in the chosen range of $\ell$ for the SKA1-Low, in the given observing frequency range, the baseline pair fraction of Type~4 dominates. This is due to a large number of redundant baseline pairs contributing to the visibility correlation in a given grid. We observe that neither bias nor the excess uncertainty  in CLTGE depends on the Type~4 baseline pairs. The  baseline pair fraction of Type~1 and Type~3 contribute only 1\% and contribution of the other baseline pair types is negligible. Given that these are the baseline-pairs types that contributes to the bias and excess uncertainty, the baseline configuration of the SKA1-Low makes it an ideal instrument for such observation.

In Paper~III we have presented estimation of bias and uncertainty in CLTGE for $\Delta \nu=0$ in Band~3 of uGMRT. We found that the standard deviation of the real and imaginary parts of the residual gain errors differ by $4\%$ in a good observation day. In this work, we consider that the real and imaginary parts of the residual gain and residual bandpass errors have same statistical characteristics, i.e.  including their standard deviations and time- and frequency-correlations. We denote the standard deviation of the residual gain error ( real or imaginary part) as $\sigma_{\delta}$ and that for the residual bandpass (real or imaginary part) as $\sigma_b$. The function $\xi(\Delta \nu)$ (same for real and imaginary) is plotted for $\nu_{\rm corr} = 2$ MHz in the right panel of Figure~\ref{fig:Cl_FG_HI} for comparison ( see footnote 1). The uncalibrated visibilities have correlated gains from the instrument and ionosphere. As a part of the calibration process, the  time-dependence of the gains are estimated.  However, the time-correlation remains in the residual gains at time-scales smaller than the time-scales at which the time-dependent gain solutions are estimated.
Since we use here  $T_{int} = 1$ sec, the least possible solution interval can be $2$ sec allowing for Nyquist sampling. Hence, any time-correlation in the residual gains, at  time-scales $< 2\ T_{int} $  remains.  In all further calculations we choose $T_{\rm corr} = 2$ sec.
We consider $N_h = 8$ hours of observations per day with $N_d$ days of observations giving rise to a total of  $T_{\rm obs} = N_h \times N_d$ hours observation time.
In the next section, we use the  estimates of baseline pair fractions, $N_G$ and $N_B$  along with various values for the parameters $\sigma_{\delta}, \sigma_b, \nu_{\rm corr}$ and $T_{\rm obs}$ to find the effect of bias and excess uncertainty in the CLTGE.

\section{Results and Discussions}
\label{sec:result}

\begin{figure*}
    \centering
    \includegraphics[width=0.95\textwidth]{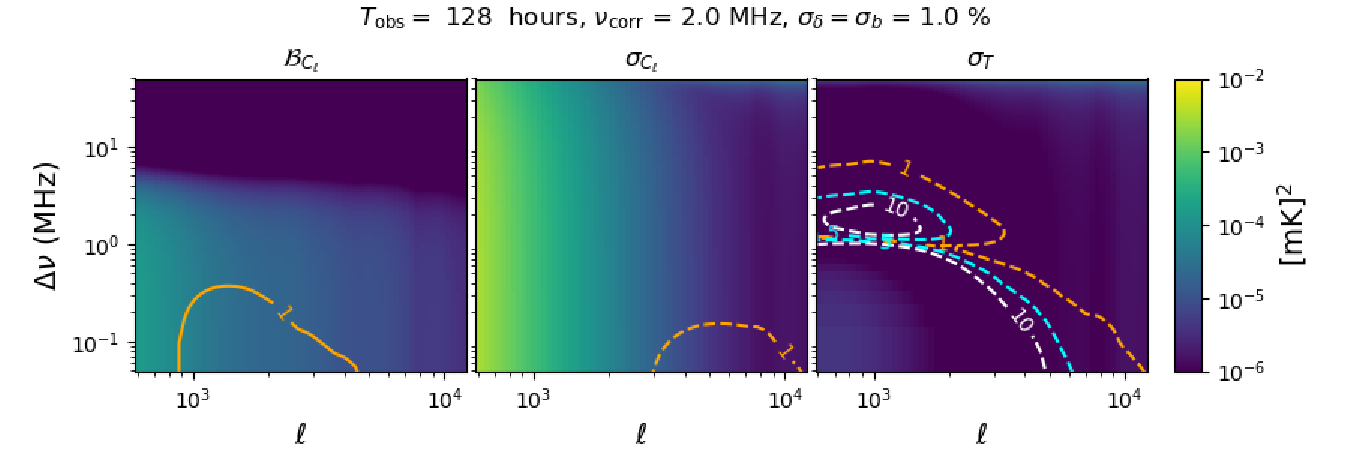}
    \caption{Bias $\mathcal{B}_{C_{\ell}}$ (left panel) and  the uncertainty $\sigma_{\mathcal{C}_{\ell}}$ (middle panel) for an observation by the SKA1-Low is shown in colormap with certain residual gain and bandpass errors as a function of angular multipole $\ell$ and frequency-separation $\Delta \nu$. The right panel show the uncertainty $\sigma_{T}$ corresponding to thermal noise only.  The colorbar  at right shows the colormap values for all three panels in [mk]$^{2}$.  In this example we have chosen a total observing time of  $128$ hours, the residual bandpass errors are considered to have a $\nu_{\rm corr} = 2 $ MHz. Both the residual gain and bandpass errors are considered to have a standard deviation of $0.01$, that is the calibration accuracy is $1\%$.  The sold contours correspond to the ratio $\mathcal{R}_{B}$ of the  fiducial \HI signal to the  $\mathcal{B}_{C_{\ell}}$, whereas dashed contours correspond to the ratio $\mathcal{R}_{\sigma}$ of fiducial \HI signal to the  uncertainty. We have shown contours corresponding to $1$ (orange), $5$ (cyan) and $10$ (white).}
    \label{fig:Fig_res1}
\end{figure*}

\begin{figure*}
    \centering
    \includegraphics[width=0.95\textwidth]{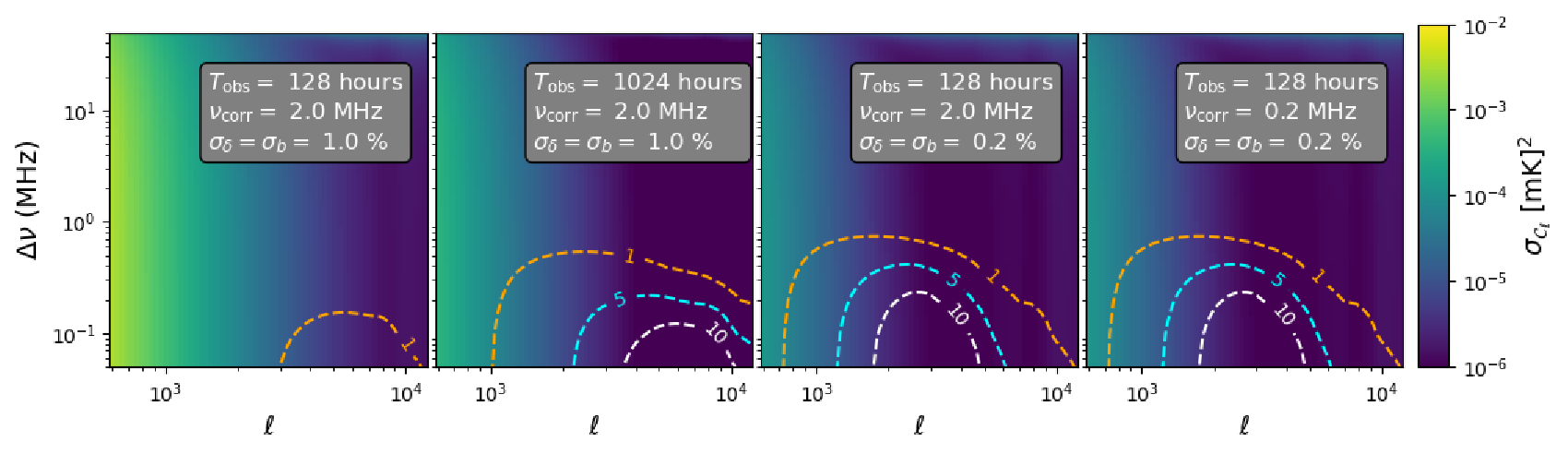}
    \caption{The uncertainty $\sigma_{\mathcal{C}_{\ell}}$  for  observations with the SKA1-Low with different observation time $T_{obs}$,  residual gain and bandpass errors $\sigma_{\delta}$ and $\sigma_{b}$ and bandpass correlation $\nu_{\rm corr}$ is shown in colormap as a function of angular multipole $\ell$ and frequency-separation $\Delta \nu$.   The dashed contours corresponds to $\mathcal{R}_{\sigma}$, the ratio of fiducial \HI signal to the  uncertainty, with colors representing $1$ (orange), $5$ (cyan) and $10$ (white).}
    \label{fig:Fig_res2}
\end{figure*}
In this section, we present our calculation of expected bias and excess uncertainty in  CLTGE in presence of residual gain and bandpass errors for various possible parameter values. \cite{2020MNRAS.494.4043M} estimate MAPS for $\Delta \nu = 0$ MHz and show its detection significance for various $\ell$ and $\nu$. Their result suggests, in presence of only thermal uncertainties, $\sigma_T$, $8-\sigma$ detection is possible with the SKA1-Low at around $z\sim 8$ for $T_{\rm obs} = 128$ hours. Since presence of gain and bandpass errors are expected to increase both the bias  $\mathcal{B_{C_{\ell}}}$ and uncertainty $\sigma_{C_{\ell}}$, we  choose a minimum observing time of $T_{\rm obs} = 128$ hours in our analysis. As it is already mentioned that we consider $8$ hours of observation per day (i.e, $N_h=8$), this would require a total of $T_{\rm obs}/N_h = N_d = 16$ observing days. We first calculate  $\mathcal{B}_{C_{\ell}}$ and uncertainty with $T_{\rm obs} = 128$ hours, $\sigma_{C_{\ell}}$ with $\nu_{\rm corr} = 2$ MHz and $\sigma_{\delta} = \sigma_{b} = 1$ \%. The calibration accuracies $\sigma_{\delta}$ and $\sigma_b$ are chosen to have optimistic values from our experience with observations with the uGMRT (see e.g. PAPER-III where the these values are $\sim 3\%$).
Figure~\ref{fig:Fig_res1} show the bias $\mathcal{B}_{C_{\ell}}$ (left panel), total uncertainty $\sigma_{C_{\ell}}$(middle panel) and thermal uncertainty $\sigma_{T}$ (right panel) using colormaps,  the colorbar at the right most part of the figure. The contour labels in the left and middle panels show the ratios $\mathcal{R}_B$ and  $\mathcal{R}_{\sigma}$  of the MAPS from the fiducial \HI signal to  the bias (solid orange line) and total uncertainty(dashed orange line) respectively. In both cases,  only the region under the orange contours have these ratios more than unity. Contours in the right most panel show the ratio of the MAPS from the fiducial \HI signal to $\sigma_{T}$, where the three contours correspond to unity (orange), five (cyan) and ten (white). This suggests, in absence of residual gain and bandpass errors, $128$ hours observation with the SKA1-Low is adequate to detect the \HI MAPS over the same $\ell$ regions as discussed in \cite{2020MNRAS.494.4043M}. We also find, that, in this condition, the signal detection is more feasible at shorter baselines and lower $\Delta \nu$ values.
 Clearly, in presence of gain error, the uncertainty increases and detection significance falls drastically. From an observational perspective, the excess uncertainty is often called as  systematics. In an observation, the measured uncertainty is essentially the total uncertainty plotted here. For an unbiased observation, hence, one would like to improve on the detection significance over the total uncertainty for a successful detection of the signal. We see that in our framework, both the bias as well as the excess uncertainty decreases with the number of observing days. The effect of gain errors can also by reduced, in principle, by increasing the calibration  accuracy and hence decreasing $\sigma_{\delta}$ and $\sigma_b$ and by estimating and mitigating the frequency-correlation in the gain error, thereby reducing $\nu_{\rm corr}$. 
 \begin{figure*}
    \centering
    \includegraphics[width=0.95\textwidth]{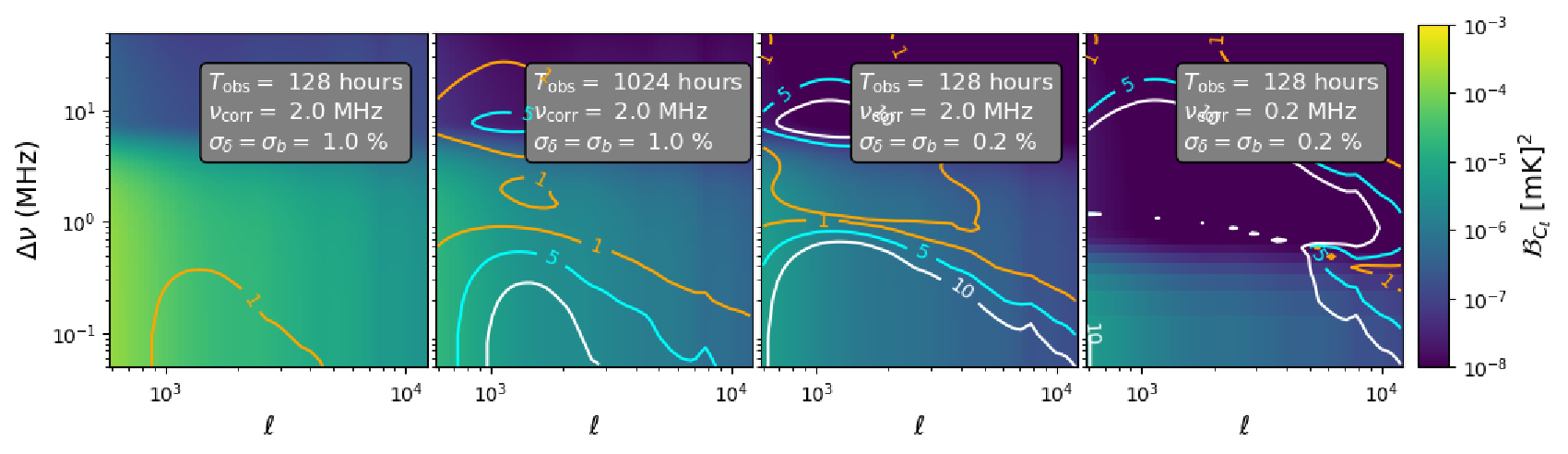}
    \caption{The bias $\mathcal{B}_{C_{\ell}}$   as a function of angular multipole $\ell$ and frequency-separation $\Delta \nu$.   The solid  contours corresponds to $\mathcal{R}_{\mathcal{B}}$, the ratio of fiducial \HI signal to the  bias, with colors representing $1$ (orange), $5$ (cyan) and $10$ (white).  The parameter values in each panel are same as that in Figure~\ref{fig:Fig_res2}.}
    \label{fig:Fig_res3}
\end{figure*}

\begin{figure*}
    \centering
    \includegraphics[width=0.95\textwidth]{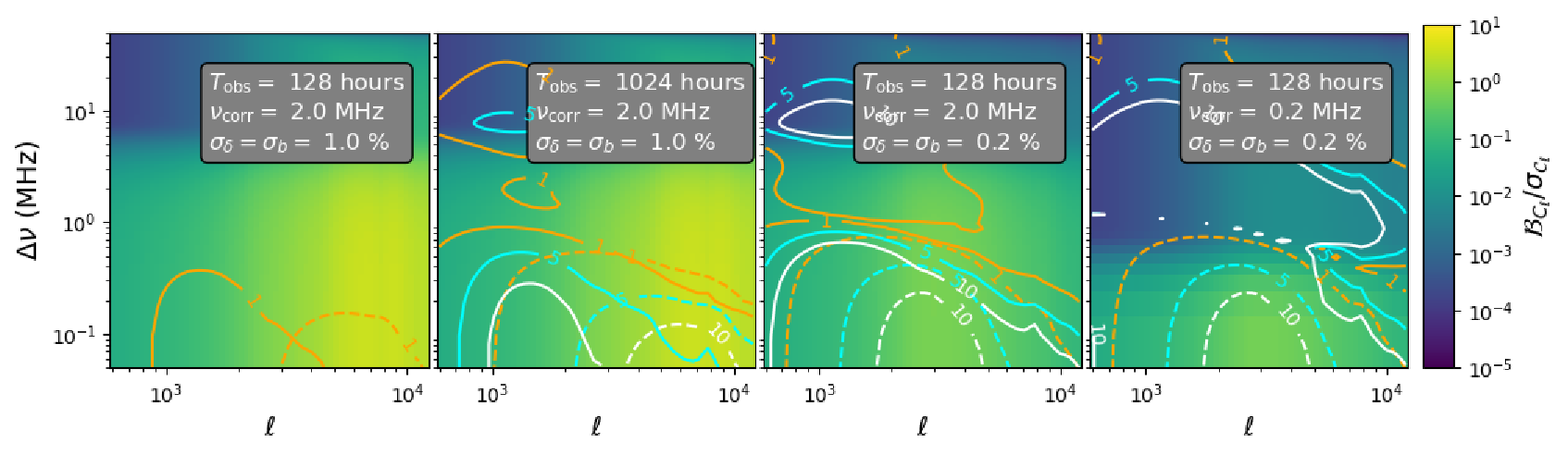}
    \caption{The bias $\mathcal{B}_{\mathcal{C}_{\ell}}$  to uncertainty $\sigma_{\mathcal{C}_{\ell}}$ ratio  as a function of angular multipole $\ell$ and frequency-separation $\Delta \nu$.   The solid  and dashed contours corresponds to  $\mathcal{R}_{\mathcal{B}}$ and $\mathcal{R}_{\sigma}$  with colors representing $1$ (orange), $5$ (cyan) and $10$ (white). The parameter values in each panel are same as that in Figure~\ref{fig:Fig_res2}.}
    \label{fig:Fig_res4}
\end{figure*}

\begin{figure*}
    \centering
    \includegraphics[width=0.95\textwidth]{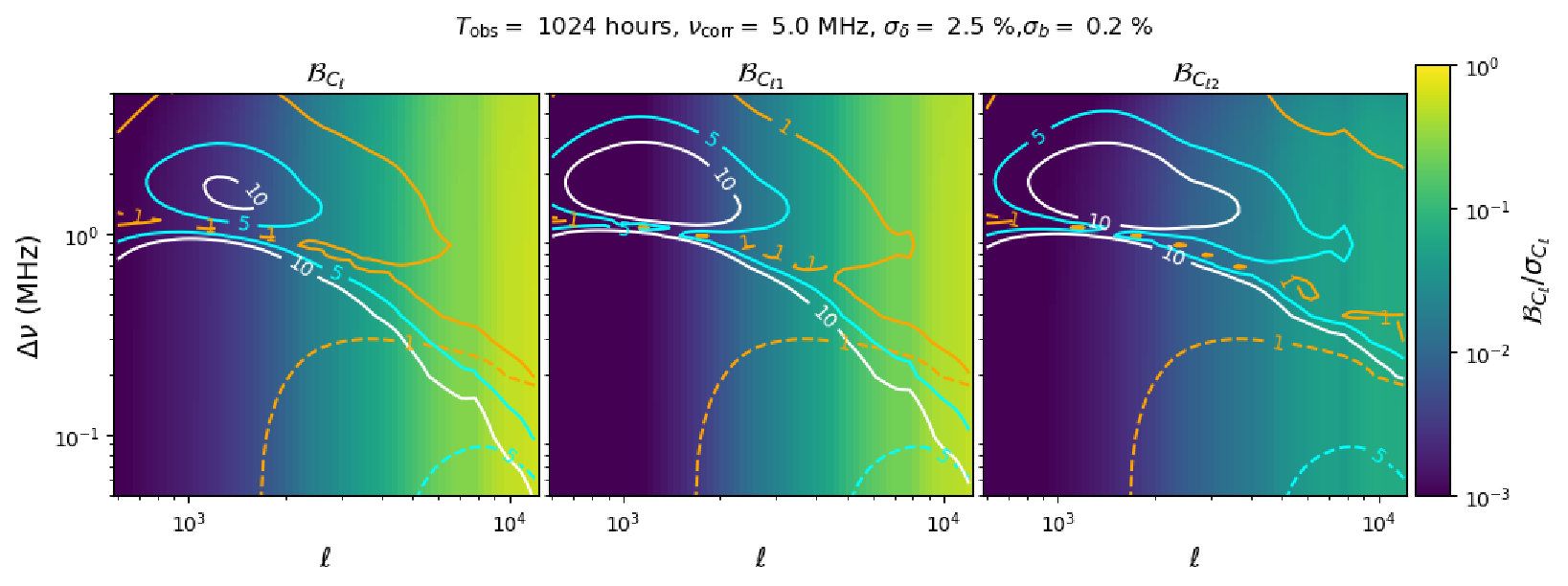}
    \caption{We compare the $\mathcal{B}_{\mathcal{C}_{\ell}}$  to uncertainty $\sigma_{\mathcal{C}_{\ell}}$ ratio for the total bias (left panel), $\mathcal{B}_{C_{\ell 1}}$ (middle) and $\mathcal{B}_{C_{\ell 2}}$ (right) calculated for $T_{\rm obs} = 1024$ hours, $\nu_{\rm corr} =  5$ MHz, $\sigma_{\delta} = 0.002$ and $\sigma_b = 0.025$. The solid and dashed contours are for $\mathcal{R}_{\mathcal{B}}$ and $\mathcal{R}_{\sigma}$ as before.}
    \label{fig:Fig_res5}
\end{figure*}

Figure~\ref{fig:Fig_res2} show the  $\sigma_{C_{\ell}}$ estimates for various values of $T_{\rm obs}, \nu_{\rm corr}, \sigma_{\delta}$ and  $\sigma_{b}$ with colormaps. The contour levels represent detection significance $\mathcal{R}_B$ as ratio of the MAPS from the fiducial \HI signal to           $\sigma_{C_{\ell}}$. These contours are marked with dashed lines and the  values are unity (orange), five (cyan) and ten (white). Henceforth, we refer to  these contours as CS. In all cases, the detection significance is higher in a limited region in the $\ell$ and limited to lower values of  $\Delta \nu$. We observe, increasing the observation time to $1024$ hours keeping other parameters fixed (first two panels) as in Figure~\ref{fig:Fig_res1}, increase the detection significance significantly.  However, the significance is now systematically higher at higher values of $\ell$. This is because the factor $N_G$ increases at higher $\ell$ and reduces the values of $\sigma_{C_{\ell}}$. At $\ell>10,000$ the detection significance decreases again, as here the signal itself is lower (see Figure~\ref{fig:Cl_FG_HI}). Note that, at larger $\ell$  the values of $N_B$ decreases, which  also decrease the detection significance.  However, contribution of the later to the total  $\sigma_{C_{\ell}}$ is rather low at higher $\ell$. We can compare the first to the third panel of the same plot to see the effect of calibration accuracy. With five times better calibration accuracy, a significant detection is possible even with only $128$ hours of observations. Noticeably,  the detection range in $\ell$ shifts towards the left, this is expected; as for infinite calibration accuracy, that is for the case of thermal uncertainty, the detection is better possible at lowest $\ell$ values here. Since, we have also reduced the value of $\sigma_b$, which results in a higher detection significance to relatively large $\Delta \nu$ values. Finally in the right most panel of the same figure, we consider the case of $\nu_{\rm corr} = 0.2$ MHz. Interestingly, change in the bandpass correlation by a factor of ten does not change the detection significance. The $\sigma_{C_{\ell}}$ is less sensitive to the bandpass correlation.

The detection significance discussed so far, in absence of bias, provides methods to choose the optimal strategy for signal detection, where, we tune our telescope and observational parameters to increase the detection significance. In presence of bias, if one is not aware of it, a rather dubious condition may arise. In an extreme case, one can find the estimated value of a signal is significantly higher than its estimated uncertainty and assert a successful signal detection. Now, if the estimated value may have more bias than the estimated uncertainty, the estimated signal is not significant in reality. Hence, we investigate the bias introduced in the measurements through the residual gain and bandpass errors. 

Figure~\ref{fig:Fig_res3} show the  $\mathcal{B}_{C_{\ell}}$ estimates for various values of $T_{\rm obs}, \nu_{\rm corr}, \sigma_{\delta}$ and  $\sigma_{b}$ with colormaps. As expected, keeping other parameters fixed (left-most two panels), high observation time reduces the bias.
The contour levels represent the  ratio $\mathcal{R}_B$ of the MAPS from the fiducial \HI signal to $\mathcal{B}_{C_{\ell}}$.  These contours are marked with solid  lines and the  values are unity (orange), five (cyan) and ten (white). Henceforth, we refer to  these contours as CB.  We observe that the effect of bias in the MAPS measurements can be more tamed by either increasing the calibration accuracy or decreasing the frequency-correlation in the residual bandpass errors. Later, has more impact on bias than the uncertainty of MAPS as the second term in eqn~\ref{eq:bias} is directly proportional to the frequency-correlation function $\xi(\Delta \nu)$.  These suggest, aiming to a higher calibration accuracy is a better strategy than increase of the observation time. 

In Figure~\ref{fig:Fig_res4},  we compare the bias and total uncertainty for  various values of $T_{\rm obs}, \nu_{\rm corr}, \sigma_{\delta}$ and  $\sigma_{b}$. We show both the contours CS and CB here. Observationally, for successful detection,  there needs to be a  range in $\ell$ and $\Delta \nu$ values where  both the $\mathcal{R}_B$ and $\mathcal{R}_{\sigma}$ are  higher than a threshold. We see, that  for $T_{\rm obs} = 1024$  hours (second panel), the region in $\ell$ and $\Delta \nu$, where both $\mathcal{R}_B > 5 $ and $\mathcal{R}_{\sigma} > 5$  is limited. Hence, for   $\nu_{\rm corr} = 2$ MHz and $\sigma_{\delta} = \sigma_{b} = 1$ \% even with large  observing time, a statistically significant detection of the MAPS is difficult. On the other hand, improving the calibration accuracy does result in significant detection (see two right panels) even with only $128$ hours of observation. Optimistically, we may estimate the bias and subtract it from the observed estimate of CLTGE.  This is particularly necessary when the bias is higher than the uncertainty in the signal.  A reliable removal of the bias, by such a method, requires us to estimate it to the accuracy defined by the uncertainty. In practice, such an estimate may be difficult \citep{2024MNRAS.534L..30A}. Colormap in Figure~\ref{fig:Fig_res4} represents the ratio of the $\mathcal{B}_{C_{\ell}}$  to $\sigma_{C_{\ell}}$. We observe that in the left two panels, where the calibration accuracy is less, for a significant region in the $\ell$ and $\Delta \nu$ values the $\mathcal{B}_{C_{\ell}} > \sigma_{C_{\ell}}$. Again, as expected, improving calibration accuracy may reduce the need to estimate the bias.

\begin{figure*}
    \centering
    \includegraphics[width=0.45\textwidth]{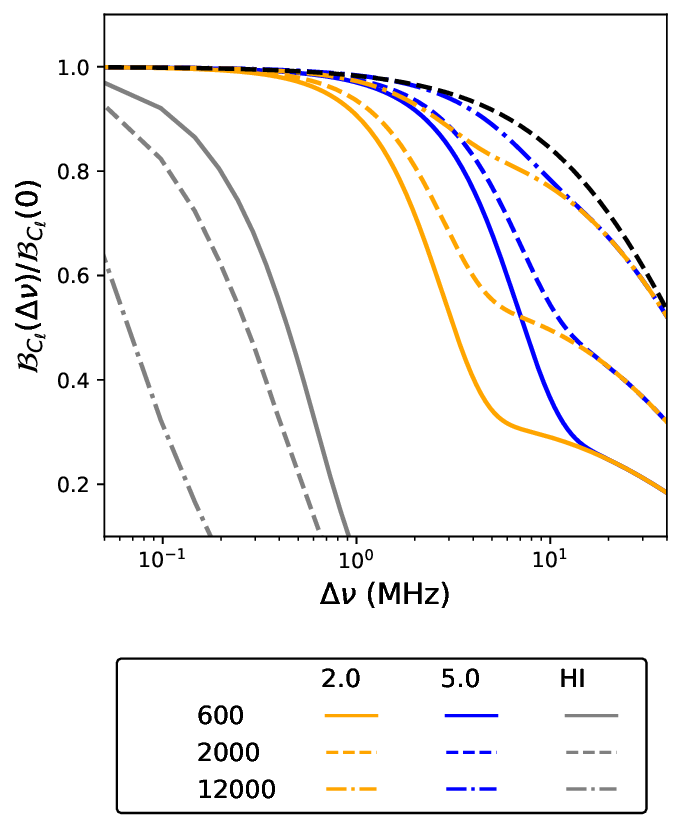}
    \includegraphics[width=0.45\textwidth]{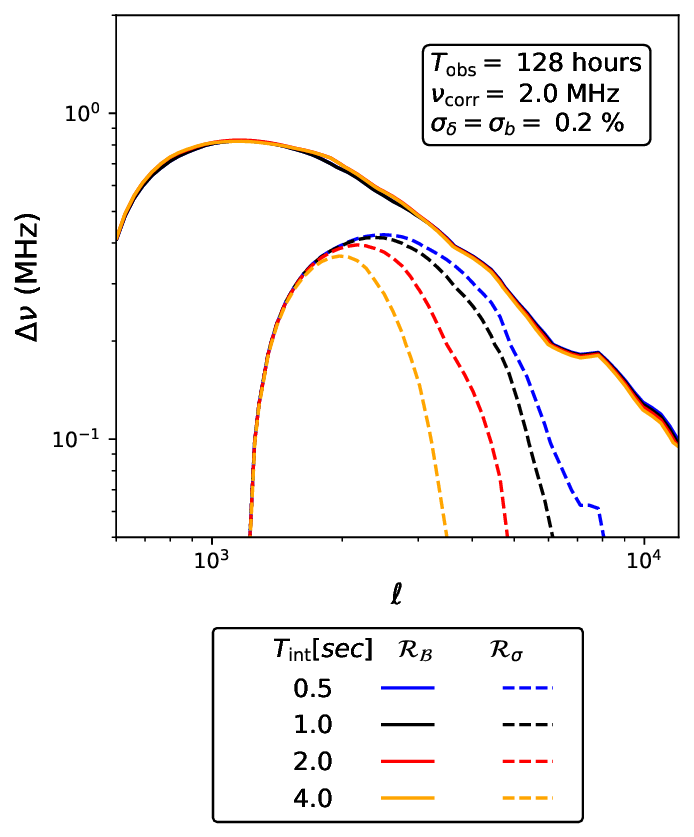}
    \caption{ Left panel: The quantity $\mathcal{B}_{C_{\ell}}(\Delta \nu) /\mathcal{B}_{C_{\ell}}(0)$ showing the frequency-de-correlation of bias in CLTGE,  as a function of frequency-separation $\Delta \nu$ for different values of $\ell$ (600, 2000, 12000) and $\nu_{\rm corr}$ ($2.0, 5.0$ MHz). The grey lines show the frequency-de-correlation  of \HI signal and the black dashed curve shows the frequency-de-correlation of foreground.
      Right panel: We plot the contours reflecting $\mathcal{R}_{\mathcal B} = 5$ and $\mathcal{R}_{\sigma} = 5$ for the integration times $0.5$ sec (blue), $1.0$ sec (black), $2.0$ sec (red) and $4.0$ sec (orange) respectively.  In this example we have chosen   $T_{\rm obs} = 128$ hours, $\nu_{\rm corr} = 2 $ MHz, $\sigma_{\delta} = \sigma_b = 0.2\%$, as in the third panel of Figure~\ref{fig:Fig_res2}, \ref{fig:Fig_res3} and \ref{fig:Fig_res4}.  The solid and dashed contours corresponds to the ratio $\mathcal{R}_{B}$ and  $\mathcal{R}_{\sigma}$ respectively.}
    \label{fig:BdnuTobs}
\end{figure*}

As discussed earlier, and also shown in Figure~\ref{fig:Cl_FG_HI}, the foreground de-correlates much slower with frequency-separation compared to the \HI in the  MAPS. Hence, in principle, one should be able to estimate the foreground from the MAPS with  large frequency-separation, extrapolate it at the smaller frequency-separation and hence subtract the foreground. Elahi et al. (2023) \cite{2023MNRAS.525.3439E} implemented this method in their CLTGE estimator and have successfully  subtracted the foreground. They demonstrate with simulation that, in absence of residual gain- and bandpass-errors, this method of foreground subtraction has a potential to recover the 21-cm power spectrum $P_{HI}(k_{\perp}, k_{||})$, from within the foreground wedge \citep{2010ApJ...724..526D}. Here we note that the $\Delta \nu$ dependence in the first term $\mathcal{B}_{C_{\ell 1}}$ of the  expression for $\mathcal{B}_{C_{\ell}}$  in eqn~\ref{eq:bias} 
originates from that in the foreground MAPS. Hence, the technique discussed in \cite{2023MNRAS.525.3439E} is expected to mitigate this part of the bias at least partially. This opens up the possibility of an unbiased detection, even in presence of significant time-dependent residual gain error, as long as effect of the second term in bias $\mathcal{B}_{C_{\ell 2}}$ is subdominant. Such a scenario is  demonstrated in Figure~\ref{fig:Fig_res5}, where we have plotted the bias to uncertainty ratio with colormap for  $\mathcal{B}_{C_{\ell}}$ (left),  $\mathcal{B}_{C_{\ell 1}}$ (middle) and $\mathcal{B}_{C_{\ell 2}}$ (right) along with the CS and CB contours. Note that, here we have chosen a relatively higher value for $\sigma_{\delta} = 2.5$   \% and a frequency-correlation $\Delta \nu = 5$ MHz, whereas the $\sigma_b = 0.2$ \%. The result is shown for $1024$ hours of observations.  Assuming that the above method of foreground subtraction is practically applicable, the part of bias given in  $\mathcal{B}_{C_{\ell 1}}$ can be subtracted,  the remaining bias to uncertainty  $\mathcal{B}_{C_{\ell 2}}/\sigma_{C_{\ell}} < 1$ for  all useful range of $\ell$ and $\Delta \nu$ (right panel).  In left panel of Figure~\ref{fig:BdnuTobs} frequency-de-correlation of the bias $\mathcal{B}_{C_{\ell}}(\Delta \nu) /\mathcal{B}_{C_{\ell}}(0)$ is plotted as a function of $\Delta \nu$  for $\ell$ (600, 2000, 12000) and $\nu_{\rm corr}$ ($2.0, 5.0$ MHz) with other parameters kept as in Figure~\ref{fig:Fig_res5}. Frequency-de-correlation of the foreground and the expected \HI signal $D(\Delta \nu)$ are shown with black dashed line and grey lines respectively. Note that, $D(\Delta \nu)$, for  the foreground model used here, is independent of $\ell$. At lower $\Delta \nu$ the curves are dominated by the second term in bias and hence the bias remains correlated to a certain value of $\Delta \nu$ depending on $\nu_{\rm corr}$. At higher $\Delta \nu$, the term $\mathcal{B}_{C_{\ell 2}}$ effectively goes to zero and the $\mathcal{B}_{C_{\ell}}(\Delta \nu) /\mathcal{B}_{C_{\ell}}(0)$ is a scaled version of $D(\Delta \nu)$ for the foreground. For large enough $\Delta \nu$, when the $\mathcal{B}_{C_{\ell 1}}$ dominates, the frequency-de-correlation depends on $\ell$. In this region,  for larger $\ell$, the signal remain correlated to higher $\Delta \nu$. Note that, if with an accurate bandpass calibration a $\nu_{\rm corr} < 1$ MHz is achieved, the bias itself is small enough (see discussion with Figure~\ref{fig:Fig_res3}) to ignore.

\section{Conclusion}
\label{sec:discussion}
In this work, we present a comprehensive analysis of the effect of time- and frequency-correlated residual gain and bandpass errors in estimation of MAPS of  the redshifted 21-cm signal in the presence of bright foregrounds. The analytical framework developed in this study extends the analytical framework presented in Paper~II by incorporating the effect of frequency-dependent residual gain errors. In this work, we assume that the residual gains can be correlated in short time-scale during the observation, however, no correlation across different days of observation is considered. Further, we also assume that the gain correlation and frequency-correlation are present only in the self correlation of an antenna with a given polarization and in  real or in imaginary parts separately.
The fiducial 21-cm signal used in this work, assumes that the 21-cm signal is ergodic in frequency and we have neglected various redshift-space distortions. However, the framework described here can be used for a more detailed non-ergodic signal. The fiducial signal is used to calculate various signal-to-bias/uncertainty ratio only.  We use this framework to predict the 21-cm signal detection possibility using the SKA1-Low through the CLTGE estimator. The major findings of this work are listed below:

\begin{itemize}
    \item The bias $\mathcal{B}_{C_{\ell}}$ and excess variance $\sigma^2_{E}$ of CLTGE  depend on the residual antenna gain and bandpass errors through their  variance and correlation in time and frequency. These also depend on the configuration of the interferometer array through the various types of baseline pairs used to calculate the visibility correlation.
    \item In general, the $\mathcal{B}_{C_{\ell}}$ and $\sigma_{E}$ increases with an increase in the residual gain and bandpass errors and the time- and frequency-correlation. In an ideal case, one expects to estimate the time-correlation of gain in time- and frequency-correlation across the bandpass. However, these assessment of gains are limited by the noise in each visibility measurement, the integration time of observation and the width of individual frequency channels. One would like to reduce the integration time as well as the channel width to reduce the residual time- and frequency-correlation, however, these increase the noise in each visibility. Hence, effect of these elements cannot be neglected for any high dynamic range detection.
    \item We observe that the baseline pairs of Type 4, where all anntennae correponding to the two baselines are different, does not contribute to the bias or excess uncertainty. A simple strategy to mitigate the bias and excess uncertainty, that emerges from this calculation, is to only use baseline pairs of Type 4 for visibility correlation. However, in practice, selective use of baseline pairs in the TGE type estimators is computationally expensive. We are working on an algorithm to improve on the computational cost and the results will be communicated in a separate study. Interestingly, a telescope with baseline distribution as the SKA1-Low has most of the baseline pairs as Type 4 and hence is a well optimized instrument for such high dynamic range observation. However, even SKA1-Low have about 2 \% contribution from the baseline pairs of Type 1 and Type 3, which contributes to the bias and excess uncertainty.  
      \item The best calibration accuracy with the modern interferometers are of the order of 1 \% for the gain and bandpass. The gain correlations are usually at time-scales of a few tens of seconds. With these parameters in mind, we found that if the excess uncertainty is only considered,  signal detection is feasible at $1024$ hours of integration at a range of $\ell$ and $\Delta \nu$. Decreasing the residual gain and bandpass errors seem to have a larger effect and at smaller time-scales the signal detection is achievable. Further, the range of $\ell$ and $\Delta \nu$ values at which the signal detection seems feasible shifts to lower $\ell$ as the residual gain and bandpass errors are decreased. Further, the fiducial signal comes significantly higher than both the bias and variance at the same range of $\ell$ and $\Delta \nu$ values only when the gain and bandpass errors are reduced. Reduction of frequency-correlation in bandpass error decreases the bias, however, the excess uncertainty is almost independent of it. These suggest that we should try to tune our calibration procedure to reduce the residual gain and bandpass errors, there can be rather less emphasis on the frequency-correlation in bandpass error for the excess uncertainty.
      \item We found that a part of the bias in CLTGE  has similar frequency-correlation as that of the foreground. Hence this part of the bias can be mitigated through the foreground mitigation methodologies that exploit the difference in  frequency-dependence of the 21-cm signal with respect to the foreground. Hence, a stable bandpass with low frequency-correlation would be key for unbiased EoR signal detection.
\end{itemize}

Here we have used the integration time as $T_{\rm int} = 1$ sec for all our analysis. This is based on the fact that at larger integration time, we expect to see time-width smearing effects. Given an integration time, the best reduction of the time-dependent gains can be done at a timescale of $2 \times T_{\rm int}$ and any time-correlation at smaller time-scales will always remain in the residual gains. Increasing the integration time, though reduces the noise in each visibility measurement, increases the remaining time-correlation in the residual gain. 
This is reflected in Figure~\ref{fig:BdnuTobs}, where we plot the contours reflecting $\mathcal{R}_{\mathcal B} = 5$ and $\mathcal{R}_{\sigma} = 5$ for the integration times $0.5$ sec (blue), $1.0$ sec (black), $2.0$ sec (red) and $4.0$ sec (orange) respectively. In each case we make sure that $T_{\rm corr} = 2 \times T_{\rm int}$ and a proper value for $\sigma_N$ is used,  however, other parameters are kept fixed as the best case discussed earlier and given in the textbox within the figure.  Note that, the bias is almost independent of $T_{\rm int}$, however, with increase of $T_{\rm int}$, the region in $\ell-\Delta \nu$ where the contours are $>5$ decreases. We observe that for integration time of $0.5$ sec, the detection significance improves. However, it needs to be kept in mind that at smaller time-scales $\sigma_N$ increases and hence the residual gain errors are  expected to be more.

The quantitative results presented in this work depend on the interferometer array configuration, the model foreground,  the gain and bandpass error models and the fiducial \HI MAPS. However, the qualitative understanding from the results are rather general and should be taken with more emphasize. The residual gain and bandpass error manifests themself in all high dynamic range interferometric observations, where they affect the detection of any fainter signal in presence of a much larger flux density signal  in the same field of view. This work demonstrates that given a known  telescope and foreground characteristics, the residual gain error effects can be estimated apriori and the observation strategy can be planned accordingly.  Here, we have assumed that the different antenna in the interferometric array have similar gain and bandpass characteristics and residual gain and bandpass errors from  real/imaginary part, different polarizations and different antennae are all uncorrelated. Further, we also assume that the residual gain and bandpass errors are mostly Gaussian random and correlations higher than the second order are zero. In Paper~III, we have checked these assumptions for uGMRT Band~3 observation for the residual gain errors, where we see that for most of the antennae the assumption hold. This is expected to be the design choice for the SKA1-Low, however, one certainly needs to assess the validity of the assumptions for a particular observation, before using the methodology presented here. In fact, in Paper~III we found that leaving out a few antennae that does not have such characteristics does improve the bias and uncertainty of power spectrum estimates.
We are recently estimating the bias and uncertainty in  CLTGE from the uGMRT Band~2 and Band~3 observations and will be presented separately.

% 4. Foreground model needed.... somethign assumed... HI model simple one..

In this work, we have used a parametric foreground model from \cite{2005ApJ...625..575S}. The foreground, in reality, is not only more complex, it also varies across different lines of sight of observation. Needless to say, in practice, one should estimate the foreground MAPS from observations along the particular line of sight of the EoR experiment for assessment of the bias and uncertainty due to residual gain and bandpass errors. In fact, in Paper~III, we have used this approach in effect.

Here, we show the effect of residual gain and bandpass error in the directly measured quantity, the MAPS, through CLTGE. The TGE based power spectrum estimators use CLTGE to estimate the cylindrically and spherically average power spectra by taking a Fourier transform along the $\Delta \nu$ axis. These, then can be mapped to the theoretical predictions to better understand reionization physics. Translating the bias and excess uncertainty in CLTGE  to the bias and uncertainty of cylindrically and spherically average power spectra, however, is a little more involved. This can be done operationally  through a simulation of visibilities with expected sky and interferometric gain and bandpass errors. This approach was tried in Paper~I, but it requires significant computational resources. In this work, we have presented a much faster analytical methodology to estimate the bias and uncertainty in CLTGE estimates. We are at present working on a simulation-less approach to translate the bias and uncertainty in CLTGE to that in the cylindrically averaged power spectrum. This will be presented along with the observational results from uGMRT Band~2 and Band~3.

\section*{ACKNOWLEDGMENT}
PD acknowledges discussion with Wasim Raja about various aspects of this work. PD would also like to acknowledge the research grant MATRICS (MTR/2023/000982) by SERB, India for funding a part of this work. SC would like to thank Philip Bull for useful discussions. The support and the resources provided by ‘PARAM Shivay Facility’ under the National Supercomputing Mission, Government of India at the Indian Institute of Technology, Varanasi, are gratefully acknowledged.

\section*{DATA AVAILABILITY}
No new data were generated or analyzed in support of this research.

\bibliographystyle{JHEP}
\bibliography{references.bib}
\pagebreak
\appendix
\section{Appendix A}
%\label(sec:appendix}
Here we present the steps to calculate the bias ($\mathcal{B}_{C_{\ell}}$) and excess variance ($\sigma^2_{E}$) in the 21-cm power spectrum measurements due to residual gain errors and bandpass errors in  presence of  strong foregrounds. The calculations presented here lead to the expressions eqn~\ref{eq:bias} and eqn~\ref{eq:sigmaE}  in sec~\ref{subsec:BVAPS}.  The calculation follow similar assumptions and methodology as in  Paper II and additionally include the effect of bandpass errors. A set of further assumptions are used to simplify the calculations while including the bandpass errors. For clarity of the reading, we present all the assumptions in italics font. 

The estimator of multifrequency angular power spectrum  (MAPS \cite{2007MNRAS.378..119D} used here is the TGE based estimator CLTGE \citep{2019MNRAS.483.5694B}, that calculates the MAPS in bins of angular multipole $\ell = 2 \ \pi U$ and the frequency separation $\Delta \nu$.
The recorded visibility is expressed in terms of the `sky' visibility, residual gain $\tilde{\mathbb{G}}_i(t, \nu)$ and correlator noise $\tilde{N}_i$ in eqn~\ref{eq:measurement}. Expressing  the residual gain $ \tilde{\mathbb{G}}_i$ in the baseline $i$,  given by the antenna pairs $A$ and $B$ at time $t$,  in terms of the antenna-based gains (eqn~\ref{eq:gain}), we write
\begin{equation}
 	\tilde{\mathbb{G}}_i =\langle \tilde{g}_A(t)  \tilde{g}^*_B(t)  \rangle \langle \tilde{b}_A(\nu)  \tilde{b}^{*}_B(\nu)  \rangle = 1 + \tilde{\mathbb{G}}_i^R,
	\label{eq:GnGr}
\end{equation}
where $\tilde{\mathbb{G}}_i^R $ can be interpreted as the excess over unity in the residual gain. Here the angle brackets in the time dependent part of the gain and bandpass  represent the average over the integration time and channel width respectively. \\
{\it Assumption I: Time- and frequency-dependence in gains are uncorrelated for a given antenna or for different antennae. }
 We consider further\\
{\it Assumption II: Antenna gains from different antenna are uncorrelated,\\
Assumption III: Real and imaginary parts of the time- and frequency-dependence in gains are uncorrelated.}
In this work we deal with a power spectrum estimation which assumes preexistence of a sky visibility model presented in \cite{2005ApJ...625..575S}  Hence, \\
{\it Assumption IV:  We have a known model for the foreground sky visibilities.}
 
As the foreground models are known, we expect to be able to subtract the visibilities due to the foregrounds. Residual visibility $\tilde{V}^R_i$ from the $i^{th}$ baseline, after subtraction of the known foregrounds, is given by
\begin{equation}
    \tilde{V}^R_i = \tilde{V}^{HI}_i + \tilde{\mathbb{G}}_i^R \tilde{V}^S_i + \tilde{N}_i,
    \label{eq:VR}
\end{equation}
where $\tilde{V}^{HI}_i$ is the redshifted 21-cm signal. Note that  $\tilde{V}^R_i$ depends on time and observing frequency. Correlating the residual visibilities, in principle, gives the angular power spectrum $C_{\ell}$. Owing to the finite beam of the antenna, the nearby `sky' visibilities remain correlated within a baseline region of $1/(\pi \theta_0)$. The estimator CLTGE  first grid the visibilities in baseline grids. Visibilities that fall in the same baseline grid but from different channels are then multiplied. To avoid noise bias, here we exclude the self correlations. The visibility correlations in each grid with same frequency separation are then collected and averaged together.   The  estimate of the $C_{\ell, g}' (\Delta \nu)$ in the baseline grid $g$ and for frequency separation $\Delta \nu$ is then given as\begin{equation}
    C_{\ell, g}' = \mathcal{R} [ \langle \tilde{V}^R_i \tilde{V}_j^{R'*} \rangle_g ],
    \label{eq:Clg}
 \end{equation}
where $\mathcal{R} [\  ]$ denotes the real part of the visibility correlation, $i, j$ denote two different baselines. The average here is taken for all baseline-pairs in a grid with a certain frequency separation $\Delta \nu$ and is denoted as $\langle \ \rangle_g$. Note that we have not written the $\Delta \nu$ dependence of $C_{\ell, g}' (\Delta \nu)$ explicitly. Here we further assume the followings:\\
{\it Assumption V:  Noise in different baselines and channels are uncorrelated.\\
Assumption VI: The sky signal, antenna gains and noise have no cross correlations.}\\
Using eqn~\ref{eq:GnGr} and \ref{eq:VR} in eqn~\ref{eq:Clg}, we can write the  power spectrum estimates in each grid as
\begin{equation} 
 C_{\ell, g}' = \langle \tilde{V}_i^{HI} \tilde{V}_j^{HI*} \rangle_g +  \langle \tilde{\mathbb{G}}_i^R\tilde{\mathbb{G}}_j^{R'*}\rangle_g \langle \tilde{V}_i^S \tilde{V}_j^{S*} \rangle_g.
  \label{eq:ClgVGV}
\end{equation}
Note that the sky visibilities $\tilde{V}_i^S$ content both the redshifted 21-cm signal as well as the foreground. The first term in the right hand side is the true estimate of the 21-cm power spectrum in the grid and can be noted as $C_{\ell_{HI,g}}$. We can use the following assumptions to simplify this expression further.\\
{\it Assumption VII:  Foreground and redshifted 21-cm signals are uncorrelated.\\
Assumption VIII: We can neglect the contribution from the 21-cm signal from $\tilde{V}_i^S$ in the last term above.}\\

\subsection*{Bias}
 The bias $\mathcal{B}_{C_{\ell,g}}$ in the estimate of the angular power spectrum  in a grid can be written as
\begin{equation} 
\mathcal{B}_{C_{\ell,g}} = C_{\ell, g}'  - C_{\ell_{HI,g}} =\langle \tilde{\mathbb{G}}_i^R\tilde{\mathbb{G}}_j^{R'*}\rangle_g C_{\ell_{FG,g}},
\end{equation}
where $C_{\ell_{HI,g}}$ and $C_{\ell_{FG,g}}$ are the redshifted 21-cm and the foreground MAPS estimates in the grid. Clearly, the terms $\tilde{\mathbb{G}}_i^R$ introduces the bias in these estimates. The quantity $\langle \tilde{\mathbb{G}}_i^R\tilde{\mathbb{G}}_j^{R'*}\rangle_g $ for a particular baseline pairs $i,j$,  in general, depends on the antenna pairs in concern, the times and the frequency  channels at which  the visibilities are recorded. As mentioned in the sec~\ref{subsec:newBPF}, in each grid, the visibility correlations are contributed from five types of baseline pairs.  Here we show the calculation for the baseline pair of Type I. In this type, the pair of baselines contributing to the visibility correlation originates from the same antenna pairs but at different times and different frequency channels. The quantity $\langle \tilde{\mathbb{G}}_i^R\tilde{\mathbb{G}}_j^{R'*}\rangle_g $ for a particular baseline pairs $i,j$ constructed by antenna pairs $A,B$ can be written as
\begin{eqnarray}
& \langle \tilde{\mathbb{G}}_i^R(t,\nu)\tilde{\mathbb{G}}_j^{R'*}(t',\nu')\rangle_g  = \langle \delta_{AR} (t) \delta_{AR} (t') \rangle + \langle b_{AR} (\nu) b_{AR} (\nu') \rangle  \\ \nonumber
& + \langle  \delta_{BR} (t) \delta_{BR} (t') \rangle + \langle  b_{BR} (\nu) b_{BR} (\nu') \rangle +  \langle \delta_{AI} (t) \delta_{AI} (t') \rangle +  \langle b_{AI} (\nu) b_{AI} (\nu') \rangle \\ \nonumber
&+ \langle \delta_{BI} (t) \delta_{BI} (t') \rangle + \langle b_{BI} (\nu) b_{BI} (\nu') \rangle,
\end{eqnarray}
where we have used assumptions I, II and III. Here the suffix $_R$ and $_I$ denote the real and imaginary part of the excess over unity in residual gain or bandpass errors. If we assume that there are a total of $N_g$ number of visibility correlations in a grid with $N_{g1}$ giving the visibility pairs of Type 1, the contribution to $\langle \tilde{\mathbb{G}}_i^R(t,\nu)\tilde{\mathbb{G}}_j^{R'*}(t',\nu')\rangle_g$ from the baseline pairs of Type 1 can be written as
\begin{equation}
 \langle \tilde{\mathbb{G}}_i^R(t,\nu)\tilde{\mathbb{G}}_j^{R'*}(t',\nu')\rangle_{g1} = n_1' (\ell) \  \left [ 2\  \Sigma_2^{\delta+ }\ \chi (\ell) + 2\  \Sigma_2^{b+ }\ \xi (\Delta \nu) \right ] 
 \end{equation}
 where $n_1'=\frac{N_{g1}}{N_{g}}$. The quantities $\Sigma_2^{\delta+ }$ , $\Sigma_2^{b+ }$  are related to the standard deviation in the residual gain and bandpass errors and are defined in sec~\ref{subsec:BVAPS}. \\
 {\it Assumption IX: Statistical properties of the gains for all antennae are similar, hence the contribution from two antennae gives a factor of two in the above expression.}\\
 As the bandpass errors in a given antenna can be correlated in frequency, we write $\xi (\Delta \nu)_I =  \langle b_{I} (\nu) b_{I} (\nu') \rangle/ \sigma_{bI}^2$. \\
  {\it Assumption X: The bandpass frequency correlation function $\xi (\Delta \nu)$ are considered to be the same for real and imaginary parts of the bandpass errors, though the variances can be different.}\\
 As the gain of an antenna remain correlated in time, the effect of this gain correlation comes though the baseline pairs with at least one antenna in common. Assuming the two point correlation function of the time dependent residual gain as $\eta (\tau)$, one can consolidate this effect in the function $\xi (\ell)$ (see eqn~\ref{eq:chidef}).\\
 {\it Assumption XI: Gain correlation function $\eta $ is  the same for real and imaginary parts of the gain, though the variances can be different.}\\
{\it Assumption XII: Time duration over which  a pair of antenna provides different baselines ($T_D$) is much larger than the integration time $T_{\rm int}$}.\\
Considering contribution from the other three baseline pairs in a similar way we can write for a given day of observation for a given baseline grid
 \begin{equation}
 \langle \tilde{\mathbb{G}}_i^R(t,\nu)\tilde{\mathbb{G}}_j^{R'*}(t',\nu')\rangle_{g}= \left [ n_{13}\chi + n_{12} \right ] \Sigma_2^{\delta+} + \left [ n_{13} + n_{12} \right ]  \Sigma_2^{b+} \xi
 \end{equation}
Here $n_{13} = 2n_1' + n_3' $ and  $n_{12} = 2n_{1A}' + n_2' $. \\
{\it Assumption XIII: The gain errors do not have any long term correlation}\\
We  use  azimuthal average around a certain baseline $U$ to estimate the angular power spectrum at the multipole $\ell = 2 \pi U$. This then gives a bias in the power spectrum estimate as
 \begin{equation}
    \mathcal{B}_{C_{\ell}}  =  \left[ (n_{13}\chi + n_{12}) \Sigma_2^{\delta+} + (n_{13} + n_{12}) \Sigma_2^{b+} \xi \right]  \frac{C_{\ell}}{N_d} ,
 \end{equation}
 where $N_d$ is the number of days of observation.

\subsection*{Variance}
We first calculate the variance of the estimator in a grid, where
\begin{equation}
    \sigma^2 _{C_{\ell}}= C_{\ell, g2}'  - \langle C_{\ell, g}'  \rangle_g ^2,
 \end{equation}
 where $C_{\ell, g2}' $ is given as
 \begin{equation}
     C_{\ell, g2}'  = \frac{1}{4} \langle [ \tilde{V}_i^R \tilde{V}_j^{R'*}  +  \tilde{V}_i^{R*} \tilde{V}_j^{R'}  ]^2 \rangle_g.
 \end{equation}
The quantities $\tilde{V}_i^R$ depends on the gain errors. \\
{\it Assumption XIV: The gain errors are Gaussian random variables.}\\
 We follow a similar procedure to calculate the variance in a grid as like for the bias for $N_d$ days of observations. Since the grid size is chosen in a way that in an annulus in baseline $U$ the estimates of the power spectrum from different grids remains uncorrelated, the variance in the power spectrum estimate can be written as
 \begin{equation}
  \sigma^2 _{C_{\ell}} = \frac{1}{N_G} \sum_{g=1}^{N_G}\sigma^2 _{C_{\ell},g},
 \end{equation}
 where $N_G$ is the total number of grid points in an annulus. Writing the total number of baseline pairs to estimate the power spectrum in a given annulus as $N_B$, the analytical expression for the variance in \HI power spectrum estimates in an annulus is
\begin{eqnarray}
\sigma^2 _{C_{\ell}}&=& \sigma^2_{T} + 2\  \frac{\mathcal{B}_{C_{\ell}}^2}{N_G} + 2 \left [  \Sigma_2^{\delta + } + \Sigma_2^{b+} \right ] \frac{N_2 C_{\ell}}{N_B N_d^2} \\ \nonumber
&+& 4 \left [  (\Sigma_2^{\delta + } + \Sigma_2^{b +} )^2 +  (\Sigma_2^{\delta - } + \Sigma_2^{b -} )^2 \right ]  \frac{C_{\ell}^2}{N_G N_d^2} \\ \nonumber
 &+& \left[ (n_{13}\chi + n_{12}) \Sigma_2^{\delta-} + (n_{13} + n_{12}) \Sigma_2^{b-} \xi \right]^2  \frac{C_{\ell}^2}{N_G N_d^2}.
    \label{eq:BiasVarA}
\end{eqnarray}
 As noted in the main text, the first terms $\sigma^2_{T}$ gives the variance of the power spectrum estimator in absence of any residual gain errors with strong foreground.

\end{document}